\documentclass[twocolumn]{emulateapj}
\usepackage{graphicx}
\usepackage{float}
\usepackage{mathrsfs}	
\usepackage{amsmath}
\usepackage{color}
\usepackage{multirow}
\usepackage{subfigure}
\usepackage{longtable}
\usepackage{blindtext}
\begin{document}
\title{New high-quality strong lens candidates with deep learning \\in the Kilo Degree Survey}

\author{\mbox{R. Li\altaffilmark{1}}}
\author{\mbox{N. R. Napolitano\altaffilmark{1}}}
\author{\mbox{C. Tortora\altaffilmark{2}}}
\author{\mbox{C. Spiniello\altaffilmark{3}}}
\author{\mbox{L. V. E. Koopmans\altaffilmark{4}}}
\author{\mbox{Z. Huang\altaffilmark{1}}}
\author{\mbox{N. Roy\altaffilmark{1}}}
\author{\mbox{G. Vernardos\altaffilmark{4,5}}}
\author{\mbox{S. Chatterjee\altaffilmark{4}}}
\author{\mbox{B. Giblin\altaffilmark{6}}}
\author{\mbox{F. Getman\altaffilmark{3}}}
\author{\mbox{M. Radovich\altaffilmark{7}}}
\author{\mbox{G. Covone\altaffilmark{3,8,9}}}
\author{\mbox{K. Kuijken\altaffilmark{10}}}

\affil{\altaffilmark{1}School of Physics and Astronomy, Sun Yat-sen University, Zhuhai Campus, 2 Daxue Road, Xiangzhou District, Zhuhai, P. R. China; {\it napolitano@mail.sysu.edu.cn, lirui228@mail.sysu.edu.cn}.}
\affil{\altaffilmark{2}Osservatorio Astrofisico di Arcetri, L.go E. Fermi 5, 50125 Firenze, Italy.}
\affil{\altaffilmark{3}INAF$-$Osservatorio Astronomico di Capodimonte, Salita Moiariello,16,80131 Napoli, Italy.}
\affil{\altaffilmark{4}Kapteyn Astronomical Institute, University of Groningen, P.O.Box 800, 9700AV Groningen, the Netherlands.}
\affil{\altaffilmark{5}Institute of Astrophysics, Foundation for Research and Technology$-$ Hellas (FORTH), GR-70013, Heraklion, Greece.}
\affil{\altaffilmark{6}Institute for Astronomy,University of Edinburgh,Blackford Hill, Edinburgh,EH9 3HJ,UK.}
\affil{\altaffilmark{7}INAF$-$Osservatorio Astronomico di Padova, Vicolo Osservatorio 5, I-35122 Padova, Italy.}
\affil{\altaffilmark{8}Dipartimento di Fisca ``E. Pancini'', University of Naples ``Federico II'', Naples, Italy}
\affil{\altaffilmark{9}INFN, Sezione di Napoli,  Naples, Italy}
\affil{\altaffilmark{10}Leiden Observatory, Leiden University, P.O.Box 9513, 2300RA Leiden, The Netherlands}

\begin{abstract}
We report new high-quality galaxy scale strong lens candidates found in the Kilo Degree Survey data release 4 using Machine Learning. We have developed a new Convolutional Neural Network (CNN) classifier to search for gravitational arcs, following the prescription by \cite{2019MNRAS.484.3879P} and using only $r-$band images. We have applied the CNN to two ``predictive samples'': a Luminous red galaxy (LRG) and a ``bright galaxy'' (BG) sample ($r<21$). We have found 286 new high probability candidates, 133 from the LRG sample and 153 from the BG sample. We have then ranked these candidates based on a value that combines the CNN likelihood to be a lens and the human score resulting from visual inspection (P-value) and  we present here the highest 82 ranked candidates with P-values  $\ge 0.5$.  All these high-quality candidates have obvious arc or point-like features around the central red defector. Moreover, we define the best 26 objects, all with scores P-values $\ge 0.7$ as a ``golden sample" of candidates. This sample is expected to contain very few false positives and thus it is suitable for follow-up observations. The new lens candidates come partially from the the more extended footprint adopted here with respect to the previous analyses, partially from a larger predictive sample (also including the BG sample). These results show that machine learning tools are very promising to find strong lenses in large surveys and more candidates that can be found by enlarging the predictive samples beyond the standard assumption of LRGs. In the future, we plan to apply our CNN to the data from next-generation surveys such as the Large Synoptic Survey Telescope, Euclid, and the Chinese Space Station Optical Survey.
\end{abstract}

\section{Introduction}
\label{sec:intro}
Strong lensing (SL, hereafter) is the effect of deformation of images of background galaxies due to the bending of their light rays from the gravitational potential of foreground systems acting as lenses or ``deflectors'' (usually massive luminous galaxies or galaxy group/clusters). This effect, predicted by General Relativity, manifests itself as spectacular arcs or rings (the so-called Einstein rings) around massive galaxies, when the source is extended. In case of point-like objects, such as high redshift quasars, multiple images of the sources are created ({\tt mupols}, hereafter) instead.

SL is a powerful tool to gain insight on the dark matter distribution in galaxies \citep{1964MNRAS.128..307R, 1992ARA&A..30..311B, 1992grle.book.....S, 1998PhDT.........6K, 2018pgl..book.....C}. For instance, it can be used in combination with dynamical analysis to determine the total mass density profiles of the lens systems \citep[e.g.,][]{2006ApJ...649..599K,2009ApJ...703L..51K,2010ApJ...724..511A,2012ApJ...757...82B,2018MNRAS.480..431L}. In case an independent inference on the stellar mass of the deflectors is available, e.g. via stellar population analysis, SL allows also to directly measure the amount and properties of the internal dark matter of the deflectors
\citep[e.g.,][]{2006ApJ...649..599K, 2009ApJ...705.1099A, 2010ApJ...721L...1T, 2011MNRAS.417.3000S,   2012MNRAS.423.1073B, 2015ApJ...803...71S, 2018MNRAS.481..819G, 2019MNRAS.489.2049N, 2019A&A...631A..40S} 

SL can also be used to measure the Hubble constant, $H_0$, as well as %some 
other cosmological parameters \citep[e.g.,][]{1964MNRAS.128..307R, 2013ApJ...766...70S, 2019MNRAS.490..613S}. In particular, this is possible by measuring the luminosity variation of lensed quasars, and using the time delay of the occurrence of their peak luminosity, which is highly sensitive to $H_0$ and little sensitive to other parameters \citep[see e.g. the H0LiCOW project,][]{2017MNRAS.468.2590S, 2017MNRAS.465.4914B}. Combining the inference obtained by more than one lens system, it has been possible to  decreased the error on the measurement of $H_0$ to 2.4\%\ \citep{2019arXiv190704869W}. This number is likely to decrease further increasing the number of systems used to infer it. 

Additionally, SL can be used to check the gravity theory by measuring the difference between gravitational lensing mass and dynamical mass \citep[e.g.,][]{2010ApJ...708..750S, 2017ApJ...835...92C, 2018Sci...360.1342C}, and it can help to search for lower mass dark sub-structures around larger galaxies and then constrain the dark matter model \citep[e.g.,][]{2012Natur.481..341V,2017MNRAS.468.1426L, 2020MNRAS.492.3047H}.  
Finally, SL can be treated as  ``natural'' telescope to study very faint high redshift galaxies otherwise impossible to be directly detected by an artificial telescope \citep[e.g.,][]{2015ApJ...808L...4A, 2018ApJ...853..148C, 2019ApJ...881....8C, 2019MNRAS.tmp.2847R, 2019MNRAS.489.5022C}. 

The probability that a distant source is lensed to produce multiple images or arcs is very small \citep{1984ApJ...284....1T, 1992ApJ...393....3F}. For instance, \cite{2008ApJ...685...57D} estimated that the galaxy-galaxy lens  candidates rate in the SDSS spectroscopic data is $\sim0.5-1.3$\%.  
Updated predictions, based on $\Lambda$CDM cosmology, suggest that, in ground-based high-resolution large sky surveys, between 0.5 and 10 lenses per square degree can be found, depending on the source nature (e.g. distant point-like quasars or extended galaxies), depth and survey strategy (\citealt{2010MNRAS.405.2579O,collett2015}).

Thus, to collect statistical samples of lensing systems, one needs to start from a very large number of galaxies and thus to use wide-sky large surveys. 
As a matter of fact, more than 1000 new lens candidates have been found in the last three years in recent ground-based surveys \citep[e.g.,][]{2017MNRAS.472.1129P, 2019MNRAS.484.3879P, 2017MNRAS.471..167J, 2019ApJS..243...17J, 2018ApJ...856...68P, 2019A&A...632A..56K}, such as the Kilo-Degree Survey \citep{2013Msngr.154...44D},  the Hyper Suprime-Cam Subaru Strategic Program \citep[HSC,][]{2012SPIE.8446E..0ZM} and the Dark Energy Survey \citep[DES,][]{2005astro.ph.10346T}. 
So far, a few hundreds systems have also been already confirmed \citep[e.g.][]{2008ApJ...682..964B, 2012ApJ...744...41B,2012AAS...21931106T,2013ApJ...777...98S,2015ApJ...803...71S, 2016ApJ...833..264S, 2018MNRAS.480.1163S, 2019MNRAS.483.3888S,2018MNRAS.479.4345A,2019MNRAS.489.2525A, 2020MNRAS.494.3491L}. 

However, despite these large numbers, the known lenses are still far from enough, especially for studies that need large statistical samples. This is particularly important in the case of distant quasars producing four multiple images (quadruplets), which are the ideal systems for cosmography. These are unfortunately also the rarest cases, representing only the 10-20\% of the full population of {\tt mupols}. The error of $H_0$ measured from a single lensed quasar is extremely sensitive to the mass distribution of its defector. Since this error is hard to be reduced under 10\% \citep[see][]{2019arXiv191105083K}, the only way to bring further down the uncertainty is to combine the analysis on a large number of systems \citep[see e.g., H0LiCOW project,][]{2013ApJ...766...70S}. 
The {\sl conditio-sine-qua-non} is therefore to find and confirm new lenses and, in the last years, we have exploited the high quality imaging offered by the KiDS survey 
to find as many previously undiscovered gravitational lenses as possible, both arcs and {\it mupols} \citep[e.g.,][]{2017MNRAS.471.3378H, 2017MNRAS.472.1129P,2019MNRAS.484.3879P,2019MNRAS.482..807P, 2018MNRAS.480.1163S, 2019MNRAS.483.3888S, 2019A&A...632A..56K}.

However, searching for strong lenses in an very large number of galaxies is a challenging task and it will become even more challenging with the advent of  next generation sky surveys. In fact, thanks to their large survey areas and deeper limiting magnitudes, upcoming surveys will effectively increase the number of lens candidates up to $\sim 10^5$ and even further  \citep{collett2015}. For instance, the optical Large Synoptic Survey Telescope \citep[LSST;][]{2009AAS...21346007C}, which will start in 2020 and cover 18\,000 $\rm deg^2$ in the Southern Hemisphere, is expected to find up to 120\,000 lenses during its operations (\citealt{collett2015}). The Space-based telescope Euclid \citep{2018LRR....21....2A}, with a point spread function of 0.2$''$ and sky areas of 15\,000 $\rm deg^2$, will find almost 170\,000 arcs and {\it mupols} \citep{collett2015}. A comparable number of lenses will also be discovered by the Chinese Space Station Telescope \citep[CSST;][]{2018cosp...42E3821Z}, which will be launched in 2024 and it is expected to cover 17\,500 sq. deg. with a PSF $\sim0.15''$. 

Traditionally, different methods, such as spectroscopic selections \citep[e.g.,][]{2006ApJ...638..703B, 2008ApJ...682..964B}, morphological recognition \citep[e.g.,][]{2007A&A...472..341S,2016MNRAS.455.1191M} and crowd-sourcing methods \citep[e.g.,][]{2016MNRAS.455.1171M}, have been used to optimize the detection efficiency. However, these methods will not be adequate for next generation surveys, since the number of galaxies that will be observed will raise dramatically, making the manual identification and selection of lens candidates impossible. 

Currently, Machine Learning \citep[ML,][]{1986mlaa.book.....M,2014sdmm.book.....I} appears to be the only viable alternative to human visual inspection to perform the lensing search task. 
This has been already shown in a number of pioneering works that have used ML techniques to search for strong lenses in some of the most successful on-going sky surveys \citep[e.g.,][]{2015MNRAS.448.1446A,  2017MNRAS.471.3378H, 2017MNRAS.471..167J, 2019ApJS..243...17J, 2018ApJ...856...68P}. In particular, our team has already actively participating in developping machine learning-based routine to search for new SL in KiDS \citep[e.g.][]{2017MNRAS.472.1129P,2019MNRAS.484.3879P,2019MNRAS.482..807P, 2018MNRAS.480.1163S, 2019MNRAS.483.3888S, 2019A&A...632A..56K}. 
Furthermore, the Strong Gravitational Lens Finding Challenge \citep{2019A&A...625A.119M}, which has compared several lensing searching methods, also demonstrated that ML methods perform as well as human inspection or other traditional methods but with a much faster classification speed. 
The general results is that thousands of new lens candidates have been found with ML methods, quickly catching up with the total number of gravitational lenses collected from traditional methods over decades.

In this context, and preparing for the big lens finding challenge with future all-sky surveys, we have started to investigate how to improve the completeness and purity of the candidates found by machine learning algorithms. In particular, in this paper, we present a new Convolutional Neural Network (CNN) classifier to search for gravitational arcs and {\it mupols}, and applied it to the $r-$band KiDS images. 
We have followed the prescription by \citet[][P+19 hereafter]{2019MNRAS.484.3879P} and developped a CNN with the same architecture but using a different training set. Furthermore, we have applied it to a larger dataset of pre-selected galaxies (for more detail about the differences, we refer the reader to Section \ref{sec:disc}), which allowed us to increase the number of high-quality lens candidates, while recovering almost all the lens candidates found from the previous CNN of P+19.

This is a preparatory work for the upcoming KIDS data release 5 (DR5, covering the full 1350 sq. deg.), and for future programs with LSST, Euclid and CSST. The paper is organized as follows. In Section 2, we describe the adopted CNN model and how we have selected the predictive data and the training sample. In Section 3, we apply our CNN classifier to the predictive data and present the new findings. In Sections 4 and 5, we make a discussion and summarize our main conclusions.

\section{A new Convolutional Neural Network classifier for KiDS}
Convolutional Neural Networks (CNNs) are one of the most popular machine learning models. Compared to traditional neural networks, the most important feature of CNNs is the use of convolution kernels as artificial neurons, which can effectively capture local features, making the CNNs particularly suitable for image recognition, speech recognition, natural language processing, as well as some other tasks \citep{1998IEEE..86..11L}. CNNs are composed of a stack of distinct layers, such as the convolutional layers (used to extract different features of the inputs), the pooling layers (used to compress the feature maps and simplify the calculations), and the fully-connected layers (used to turn all the local feature maps into a global feature map). For more information about CNNs, we refer the reader to our previous paper \citet{2019MNRAS.482..313L}, or to the recent review from \citet{2017NC...29...1R}. In general, any good CNN model, learns from the training data, provided that this is  sufficient and suitable for the  classification, and then make predictions on the predictive data.

In this work, we used a CNN to search for gravitational lenses from a large sample of $\sim10^6$ bright galaxies and $\sim10^5$ red luminous galaxies (see Section \ref{sec:predictive_data}). This machine learning based searching method is quite recent and the best architecture to choose to optimize the SL finding is not yet understood. For this reason, we have compared the performances of different architectures, such as AlexNet \citep{2012ANIPS}, ResNet \citep{2015arXiv151203385H} and a more recent one named Densenet \citep{2016arXiv160806993H}, to optimize the tool for the lensing search. As result, we decided to use a ResNet model with 18 convolutional layers, which best balanced performance and speed. For instance, AlexNet required less training time, but returned a lower precision than ResNet, while the DenseNet showed the opposite behaviour. Also, deeper ResNets (e.g. 34 or 50 layers) are expected to have better performances, but are more time consuming.

The same choice was already made in P+19, with which we share the core part of the classifier, coming from the same open-source code keras-resnet\footnote{https://github.com/raghakot/keras-resnet}. Therefore, there are no architectural differences between our classifier and that presented in P+19. Furthermore, the classifiers are both built with Keras\footnote{https://github.com/keras-team/keras}, and run on the back-end of  TensorFlow\footnote{https://github.com/tensorflow/tensorflow}. Despite the similarities between the 
%new classifier and that of P+19
%NRN: check this out
two CNNs, the new classifier has been able to find more candidates. As we will explain in the following, this is mainly because of the different training sample used to train the CNN and of the different predictive data on which we applied it.

\subsection{The predictive data}
\label{sec:predictive_data}
Predictive data are systems over which the trained CNN can return a probability, $p_{\rm cnn}$, (i.e. make a prediction) to be a real lenses (true positive). In principle, all targets detected in a survey can be part of the predictive sample. However, it makes no sense to feed the CNN with stars, quasars, low-redshift dwarf galaxies or other very fainter galaxies, because they cannot act as gravitational lenses. Thus, a pre-selection can be done a-priori to help reduce the computation time and potential contamination. Since the SL cross-section is larger for massive galaxies \citep[see e.g. ][]{2010MNRAS.405.2579O}, a standard approach consists in using only the brighter and more massive systems as the predictive data.
 
To build our predictive data, we used the 1006 publicly available tiles from the latest KiDS data release, KiDS-DR4. 
This contains a multi-band optical catalog extracted from images in four optical bands ($u$, $g$, $r$, and $i$).   Here we used only the $r$ band observations since they have the best seeing with a median full width at half-maximum (FWHM) of $\sim0.7''$ \citep{2019A&A...625A...2K}. In an upcoming paper, we will further improve the results by exploiting $g, r, i$ color-composite images, hence using information on the colours for both the lens and the arcs/{\it mupols},  together with arc-morphology and image positions. However, this has to be done in a careful way, and only if a proper training sample, well describing the population of real galaxies and their color distribution, is available. In fact, the lens colours can be contaminated by the presence of the source and, thus, not matching with a simple color-cut designed to select LRGs.

The total number of detected sources in the publicly available KiDS-DR4 catalog is  $\sim$120 million, of which more than 60 million are galaxies with high-quality photo-z obtained with BPZ code\footnote{http://www.stsci.edu/~dcoe/BPZ/}  \citep[see][]{2019A&A...625A...2K}. Among these, more than 5 million have also structural parameters from seeing convolved 2D single  S{\'e}rsic model \citep[Roy et al. in preparation, see also][for the analysis of KiDS-DR2]{2018MNRAS.480.1057R}

In this work, we applied our CNN classifier to two predictive datasets. The first dataset (referred as LRG sample), comprises only Luminous Red Galaxies (LRGs), which are more likely to exhibit strong lensing features, being generally more massive.  Therefore, they are commonly used as standard pre-selection sample in arc-finding searches \citep[][P+19]{2013ApJ...769...52W,2017MNRAS.472.1129P,2019MNRAS.484.3879P}. In addition, as second predictive dataset, we added a much larger sample of ``bright galaxies" (BGs, referred as BG sample), without any color cut. 
This is for two main reasons: 1) the color cuts to define LRGs are arbitrary and might not be optimal in the case of SL, where the lensed images can contaminate the colors of the lens (especially in cases where the Einstein radius is small; 2) SL can be produced by distant massive galaxies, regardless their morphology/color. 

Furthermore, the fastest GPUs allow us today to analyze a larger amount of data with almost no increase in the total computing time \citep[see, e.g.,][]{2016arXiv160304467A}. Of course, even if adding also the BGs to the predictive sample increases the chance of finding new lenses, at the same this also causes a larger contamination from false positives.

We give a description of the two predictive samples here below:
\begin{enumerate}
\item {\tt BG sample}: In the KiDS catalog, the BG sample has been chosen by: 1) selecting galaxy-like objects using the flag {\tt SG2DPHOT}=0. This flag is derived by the software {\tt 2DPHOT} \citep{2008PASP..120..681L}, which performs a star-galaxy separation in the KiDS catalog extraction process \citep[see][for KiDS-DR4]{2019A&A...625A...2K} and assigns a zero value to galaxies and values larger than zero to  point-like objects. 2) requiring the $r-$band Kron-like magnitude {\tt mag$\_$auto} (also present in KiDS catalogs and obtained by {\tt Sextractor}, \citealt{1996A&AS..117..393B}) to be $r_{auto}\leq21$. The final BG sample selected with these two criteria consists of 3\,808\,963 galaxies. 

\item{\tt LRG sample}: The LRG predictive sample is a subsample of the BG sample, where we have followed the approach from P+19, slightly adapted the low-redshift ($z < 0.4$) LRG color-magnitude selection in \cite{2001AJ....122.2267E} to include fainter and bluer sources:
\begin{equation}
\begin{split}
r_{auto}<14+c_{par}/0.3,\\
|c_{perp}|<0.2,
\end{split}
\label{eq:cuts}
\end{equation}
where
\begin{equation}
\begin{split}
c_{perp}=(r-i)-(g-r)/4.0-0.18,\\
c_{par}=0.7(g-r)+1.2[(r-i)-0.18],
\end{split}
\label{eq:colours}
\end{equation}

being $r_{auto}$ the $r$ band Kron-like magnitude as above. 
We restricted the selection to $r_{auto}\leq20$ for LRGs to match the P+19 prescription. Galaxy colors have been directly retrieved by the KiDS-DR4 catalogs from the flag {\tt COLOUR$\_$GAAP$\_ $g$\_$r} ($=g-r$) and  {\tt COLOUR$\_$GAAP$\_$r$\_$i} ($=r-i$). These colors are different from the ones used in \cite{2017MNRAS.472.1129P,2019MNRAS.484.3879P,2019MNRAS.482..807P}, which were based on Kron-like magnitudes. 
In fact, Kron-like magnitudes in other bands are not anymore listed in the KiDS catalog after KiDS-DR3 and thus they are not not publicly available for all sources in DR4.
On the other hand, the {\tt COLOUR$\_$GAAP} were measured on Gaussian-weighted apertures, which are modified per-source and per-image, so they provide seeing-independent flux estimates across different observations/bands, hence providing more unbiased colors (\citealt{2019A&A...625A...2K}, P+19). 
Using the criteria in Eqs. \ref{eq:cuts} and \ref{eq:colours} we have obtained a sample of 126\,884 LRGs. 
\end{enumerate}

For both BG and LRG sample, we extracted cutouts of $101 \times 101$ pixels, corresponding to $20 \times 20$ arcsec$^2$, centered on each of these galaxies, from the $r$ band coadded images from KiDS-DR4. The cutout sizes (corresponding to 90 kpc $\times$ 90 kpc at $z=0.3$ or 120 kpc $\times$ 120 kpc at $z=0.5$) are large enough to enclose from galaxy-sized to group/cluster-sized arcs and {\it mupols}, and also to have a sense of the environment around the lens candidates.

\subsection{The training data}
\label{sec:training}
The training data represents the dataset from which the CNN has to learn which features should be detected in the predictive dataset to allow the classification. In general, it is composed of ``true positives'', i.e. real confirmed lens systems and ``true negatives'', i.e. systems containing no detectable lensed images, but which can contain features similar to the ones of true lensing events that the CNN has to learn to exclude (e.g. blue spiral arms mimicking a lensed arc, or ring galaxies mimicking Einstein rings, see a more detailed discussion below). Moreover, the training sample needs to realistically reproduce the data quality of the predictive sample. In case this condition is not fulfilled, and the training sample does not recover the main attributes of a predictive sample, domain adaptation or transfer learning \citep[see e.g.,][]{2018arXiv181211806K} can be applied. This, for instance, will be a necessary approach in the future, when color information will be added to the CNN for the classification.

Since we do not have a large sample of real lenses in KiDS (i.e. most of the candidates from P+19 and other papers are not confirmed yet)\footnote{We note that the only possible rigorous definition of confirmed or rejected lenses comes from spectroscopic confirmation, as visual inspection does not provide a proof that a candidate lens is real.}, to build up ``true positives'', we simulated realistic arcs around a selected sample of galaxies extracted randomly from the predictive sample \citep[see e.g.][]{2017MNRAS.472.1129P}. To this purpose, we followed the description in P+19. We used a singular isothermal ellipsoid (SIE) profile plus external shear to model the deflectors and an S{\'e}rsic profile to model the light of the background sources.  The model parameters have been set as in \cite{2019MNRAS.482..807P}, where they have been demonstrated to be realistic enough. In particular, the Einstein radius, 
%derived by the SIE model and the redshift of the deflector and source, 
was set to be in the range [1, 5] arcsec, and to follow a logarithmic distribution. Additionally, the Gaussian random field accounting for the effect of the sub-halos of the deflector, and small light blocks (modelled with S{\'e}rsic profiles) reproducing the corresponding source substructures, implemented in P+19, were also added. When training the CNN classifier, we re-scaled the brightness of the arcs by the peak light of the central galaxies and normalized all images to the same range of counts, [0, 255]. We also did data-augmentation for positives (the simulated lenses) and the negatives in the training process (e.g. rotation, shifting, flipping, rescale). 

\begin{figure}
    \centering
    \includegraphics[width=8cm]{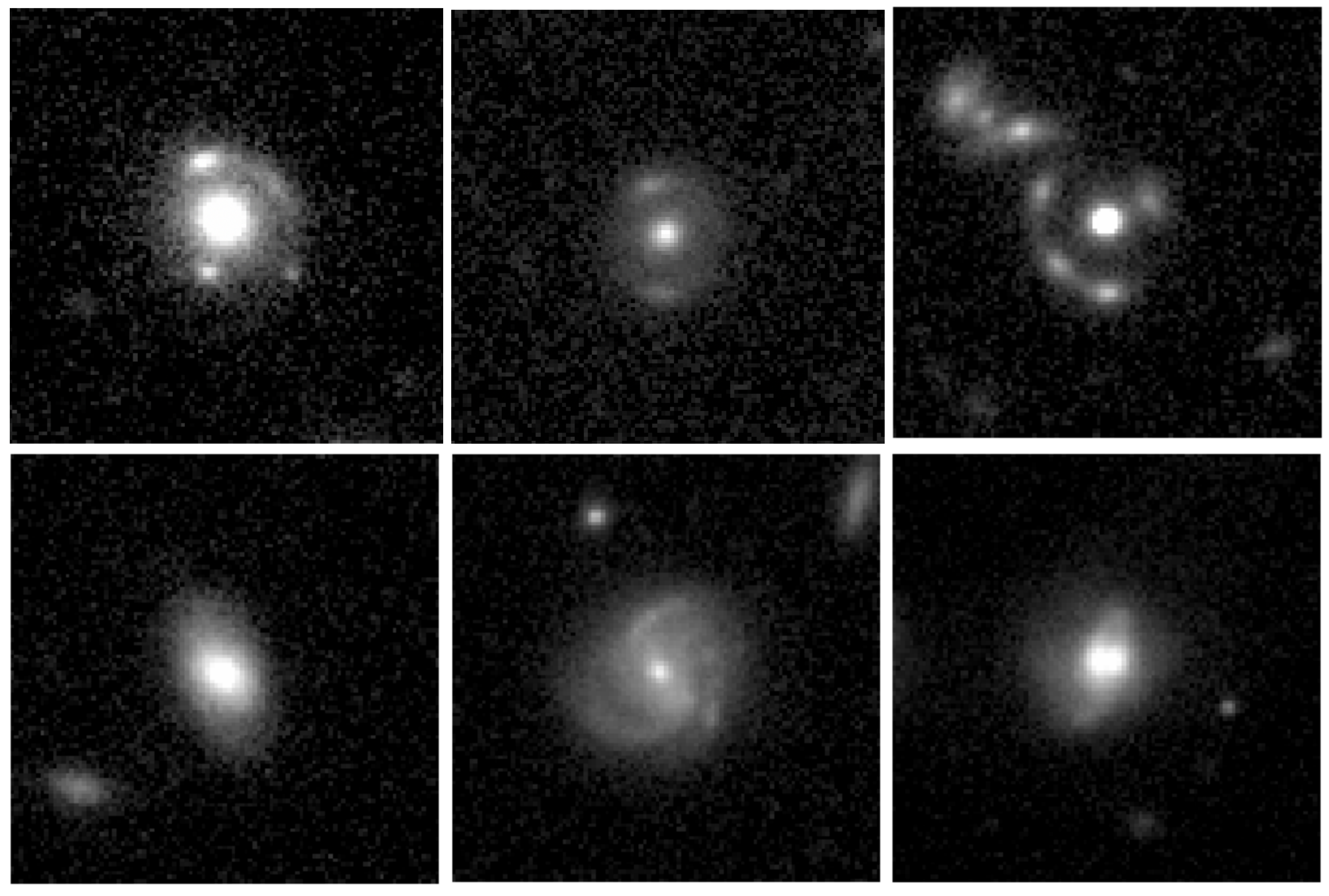}\vspace{3pt}
    \caption{Examples of the training sample. The pictures in the first row are 3 simulated lenses (`positives') produced by adding mock arcs to real LRGs. The pictures in the second row are 3 real galaxies used as `negatives'}
    \label{fig:train_sample}
\end{figure}

Thus, in summary the training data have been divided into two classes: the {\tt positives} and the {\tt negatives}. The {\tt positives} are the `true lenses', i.e. galaxies around which we know there is a (simulated) arc, that we labelled with a
[1] mark, while {\tt negatives} are the `no lens galaxies', i.e. real KiDS galaxies with no simulated arcs, and we labelled them [0] mark. Here below we describe in more details how these two classes have been constructed:

\begin{enumerate}
\item {\tt positives}: 
we have selected 11\,000 LRGs from the LRG sample, of which about half were provided by P+19 and half were selected by us via visual inspection. We then simulated 200,000 arcs and convolved them with an average point spread function (PSF) of KiDS DR4. For each arc, we randomly chose an LRG from the selected sample and added the arc to it to create a mock lens system. With this method, we built 200\,000 mock lenses, suitable to be used as {\tt positives} to train our CNN. 
\item {\tt negatives}: 
this sample is made of a total of 18\,000 real galaxies, comprising the 11\,000 LRGs that we used to simulate the {\tt positives}, 3\,000 non-lens galaxies randomly selected from KiDS-DR4, 2\,000 spiral galaxies (of which 1\,000 were provided by P+19, and another 1\,000 were selected by us through visual inspection of KiDS-DR4 images) used to train the CNN to avoid ``false positives'' produced by spiral arms, and finally 2\,000 other kind of ``false positives'' (e.g., mergers, ring galaxies, etc). In particular, for this latter class, we selected candidates that the CNNs that we built to test the different architectures (see Section~2) classified as probable lenses but that were then rejected after visual inspection.
\end{enumerate}
Fig. \ref{fig:train_sample} shows examples of the training sample. The images in the first row show 3 simulated lenses ({\tt positives}), by adding mock arcs to real LRGs. In the same figure, the second row shows 3 real galaxies used as {\tt negatives}. 

Here we stress two points. First, our assumptions do not account for correlations between the lens-galaxy properties and the lensed images in the simulating process. Although in real lenses the galaxy mass and light are correlated, we have made this choice to avoid any possible bias (e.g. assuming a specific dark-to-luminous mass ratio, e.g. from abundance matching, see \citealt{2010ApJ...710..903M}). 
Indeed, given the large uncertainties in these relations and the limited mass range of lens galaxies, the adoption of a correlation with realistic variance would have produced an almost uniform bi-variate distribution, equivalent to no-correlation. 
Second, our choice to use only LRGs to simulate real lenses in the training sample is meant to optimize the CNN predictive power primarily over this sample, but it might impact the predictive power for the bright sample. 
We stress, though, that the BGs are not expected to have attributes (e.g., luminosity, size, flattening) that differ significantly from the ones of the LRGs and that might, in some way, affect the ability of the CNN to discriminate true positives (either arcs or {\it mupols} around galaxies, see also \S\ref{sec:HQ}). 
%The most significant difference between LRG and BG samples could be the color, but it is not used in this work. 
From this point of view, the application of the CNN, trained on the LRGs, to the BGs does not justify the use of   
transfer learning \citep[see e.g.,][]{2018arXiv181211806K}. 
In principle, we could have used this approach, in case we had used multiple band information, as colour is the only distinctive parameter of the two predictive samples. Since, for the moment, we use only $r-$band we do not expect this feature to affect the CNN performance or produce overfitting. 
However, we will investigate the use of transfer learning in next analyses, when we will compare the results of CNNs trained on $r-$band only and those of CNNs trained on multi-band images.

\begin{figure*}
    \centering
    \includegraphics[width=7.35cm]{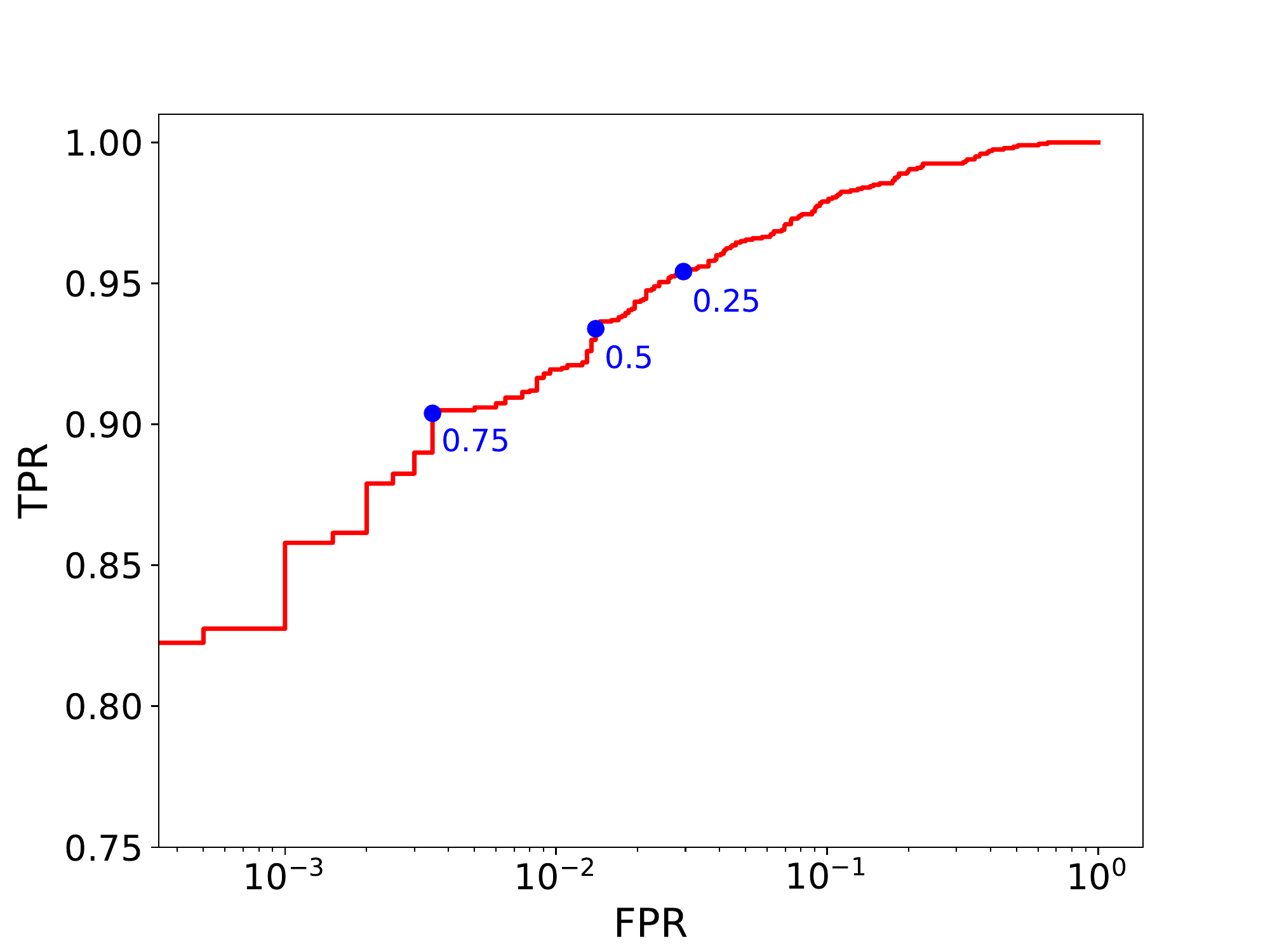}\vspace{3pt}
    \includegraphics[width=8cm]{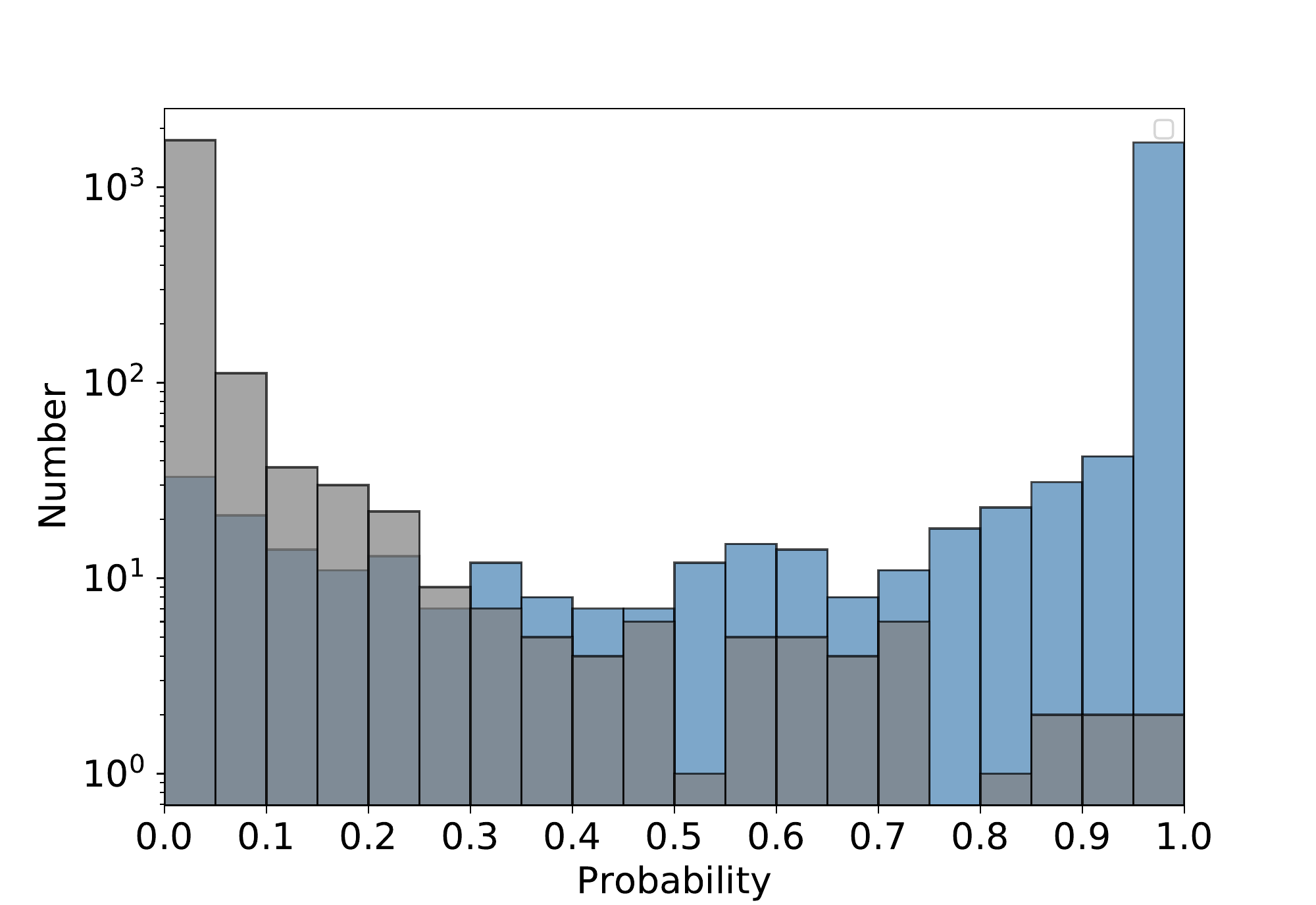}\vspace{3pt}
    \caption{{\sl Left}: the ROC curve for the CNN classifier based on 4\,000 galaxies in the testing sample. We also show the locations of 3 different values of threshold ($p_{\rm cnn}=0.25,0.5,0.75$) used to calculate the FPR and TPR. {\sl Right}: The probability distribution of the testing 
    sample. The blue histogram represents the probability distribution of the {\tt positives} while the grey histogram shows that of the {\tt negatives}.}
    \label{fig:ROC_dist}
\end{figure*}

\subsection{Testing the CNN classifier}
\label{Sec:testing}
After training, the CNN classifier has been tested on a test sample to evaluate its performances. 
The test sample was made of 2\,000 simulated lenses, following the prescription in Section \ref{sec:training}, as {\tt positives} and 2\,000 randomly selected real galaxies from the LRG sample as {\tt negatives}. We note that we only used galaxies in the LRG sample for testing, since the CNN is trained only on that.

We use the Receiver Operating Characteristic (ROC) curve to evaluate the performance of the CNN classifier (see also \citealt{2019MNRAS.482..807P}. The ROC curve is obtained by plotting true-positive rate (TPR) against false-positive rate (FPR) for different $p_{\rm cnn}$ thresholds, where TPR and FPR are defined as follows:
\begin{itemize}
\item {\tt TPR}: The fraction of {\tt positives} that also have been identified as {\tt positives} by the classifier (i.e. objects on which the classifier works properly). 
\item {\tt FPR}: The fraction of {\tt negatives} that have been wrongly classified as {\tt positives} by the classifier.
\end{itemize}

In Figure \ref{fig:ROC_dist} we show the ROC curve (left) and the probability distribution (right) of the whole testing sample (2\,000 simulated lenses and 2\,000 real non-lens galaxies both taken from the LRG sample, which is the one we use to train the CNN). The ROC curve is similar to the one in \cite{2019MNRAS.482..807P}, showing that the two CNNs perform very similarly. In the right panel of the figure, what we plot is the distribution of the output CNN probability of true {\tt positives} (i.e. lenses, in blue) vs. {\tt negatives} (i.e. non-lenses, in grey). 
The figure demonstrates that a fraction of real lenses can be lost, because they are wrongly rejected by the classifier and assigned a very low probability. We have visually inspected these cases within the testing sample, finding that the majority of missed lenses have arcs that are too faint to be recognized or that are embedded in the light of the foreground galaxy. This shows that the current CNN performs well for bright arcs while for more extreme configurations (e.g., very small Einstein radii) some improvements are still required, which we will implement in next developments.

The figure also clearly shows that for higher $p_{\rm cnn}$s, the fraction of negatives decreases. Thus, a threshold can be defined to select good candidates. In this paper, we decided to adopt $p_{\rm cnn}=0.75$, above which the fraction of negatives remains always below 6.5\%.
%above which the number of negatives remains always below 2.

\section{New lens candidates}
\label{sec:new_lenses}
The compilation of the lens candidates is based on two steps: the first step is the classification by the CNN and the second one is the visual inspection by five expert observers.  This latter step is necessary to clean the final sample from clear ``false positives'' and to add an independent score to the lenses for which the CNN has returned a high probability. This allows us to optimize the chance that a given candidate can be a real lens, as this selection process involves both artificial and human intelligence. In future large surveys (e.g., LSST, Euclid, CSST), the visual inspection by experts will represent a severe bottleneck for the process, as the number of candidates from a CNN classifier will be the order of several hundreds thousands. Hence human inspection will be doable only through some form of citizen science project \citep[see e.g.][]{2016MNRAS.455.1191M}. In alternative, we  we will need to find other automatic filtering techniques to further prune the candidates from obvious false positives and reduce the sample to visually inspect to reasonable sizes for large collaborations (e.g. of the order of several tens thousands).

\subsection{CNN probability and preliminary candidate selection}
\label{sec:pCNN}
%After the CNN training, 
After training and testing the CNN, we first applied the network to make predictions (i.e. to look for arcs) on the LRG sample. In this case, the input of the CNN is the set of 126\,884 normalized images of the LRG sample, described in Section \ref{sec:predictive_data}, while the output is the probability, $p_{\rm cnn}$, for each of them to be a lens. 

As already specified in the Section \ref{Sec:testing}, we set a threshold probability of $p_{\rm cnn}=0.75$ to define a system to be a valuable lens candidate and qualify for the visual inspection. This threshold has been set as a reasonable trade-off between the CNN probability output of true lenses and a false positive in the training run (see Fig. \ref{fig:ROC_dist}). Note that this threshold is different from the one adopted in P+19 ($p_{\rm cnn}=0.8$), but returned a similar number of potential candidates (see the discussion on Section \ref{sec:disc}).

\begin{figure}
    \centering
    \includegraphics[width=8cm]{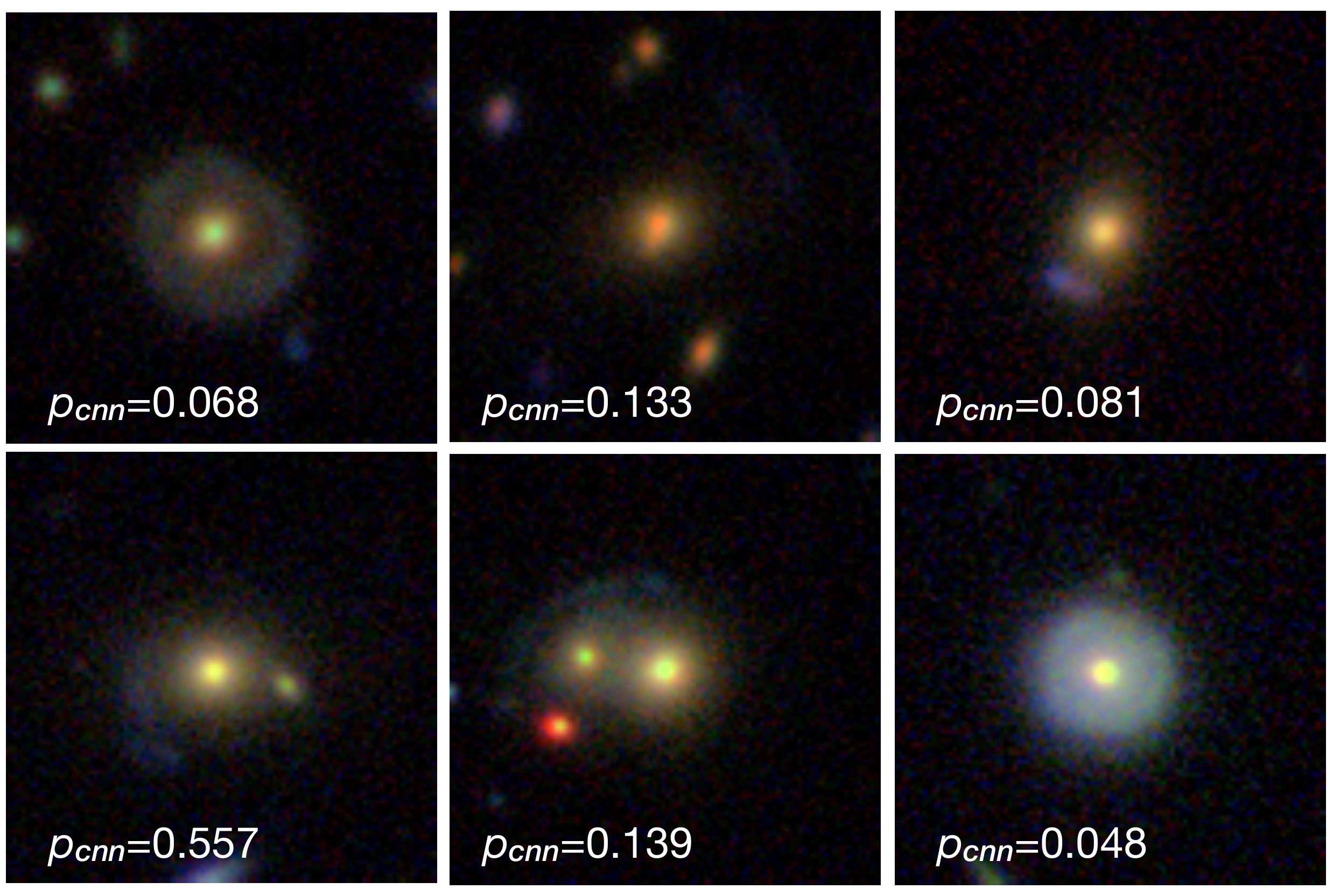}\vspace{3pt}
    \caption{The 6 lens candidates found by P+19, but missed by our CNN. $p_{\rm cnn}$ is the probabilty from our CNN classifier. The stamps ($20''\times20''$) are obtained by combining $g$, $r$, and $i$ images.}
    \label{fig:missed_sample}
\end{figure}

We have obtained 2848 candidates (2.24\% of the full LRG sample), including 54 of the 60 high-quality LinKS lenses candidates already classified by P+19, corresponding to a 90.0\% recovery rate. The 6 ``missing"  objects whose color-combined KiDS cutouts are shown in Fig.~\ref{fig:missed_sample}, have probabilities lower than the threshold we fixed. Some of them might be real lenses missed by our CNN classifier. On the other hand,  we find and present here good candidates missed by the CNN of P+19. A full comparison of the two classifiers is beyond the purpose of this paper, and will be addressed in detail in a forthcoming work. Here we note that, despite the similarities between the two classifiers, they are not identical (see the summary of the differences between them in Section \ref{sec:disc}), and, as such, they present interesting complementary aspects. This demonstrates the importance of developing independent ML classifiers (either CNNs or also other ML techniques such as support vector machines, SVM, etc.), and possibly exploiting the combined strengths of each of them to further improve the overall performances of hybrid configurations.  

We have then applied our CNN model to the full BG sample, which is however more prone to induce a larger number of false positives, since the CNN is not optimized for this sample. Moreover, the BG sample also includes slightly fainter galaxies with any color, thus also late-type systems, whose spirals could mimic arc-like lensing features.
In order to reduce the fraction of such false positives, in this case, we have set a higher (and quite conservative) probability threshold to $p_{\rm cnn}=0.98$, to accept a system as a valuable lens candidates. With this threshold, we have obtained 3\,552 lens candidates, corresponding to a fraction of 0.093\% of the BG sample.  

\begin{figure}
    \centering
    \includegraphics[width=9cm]{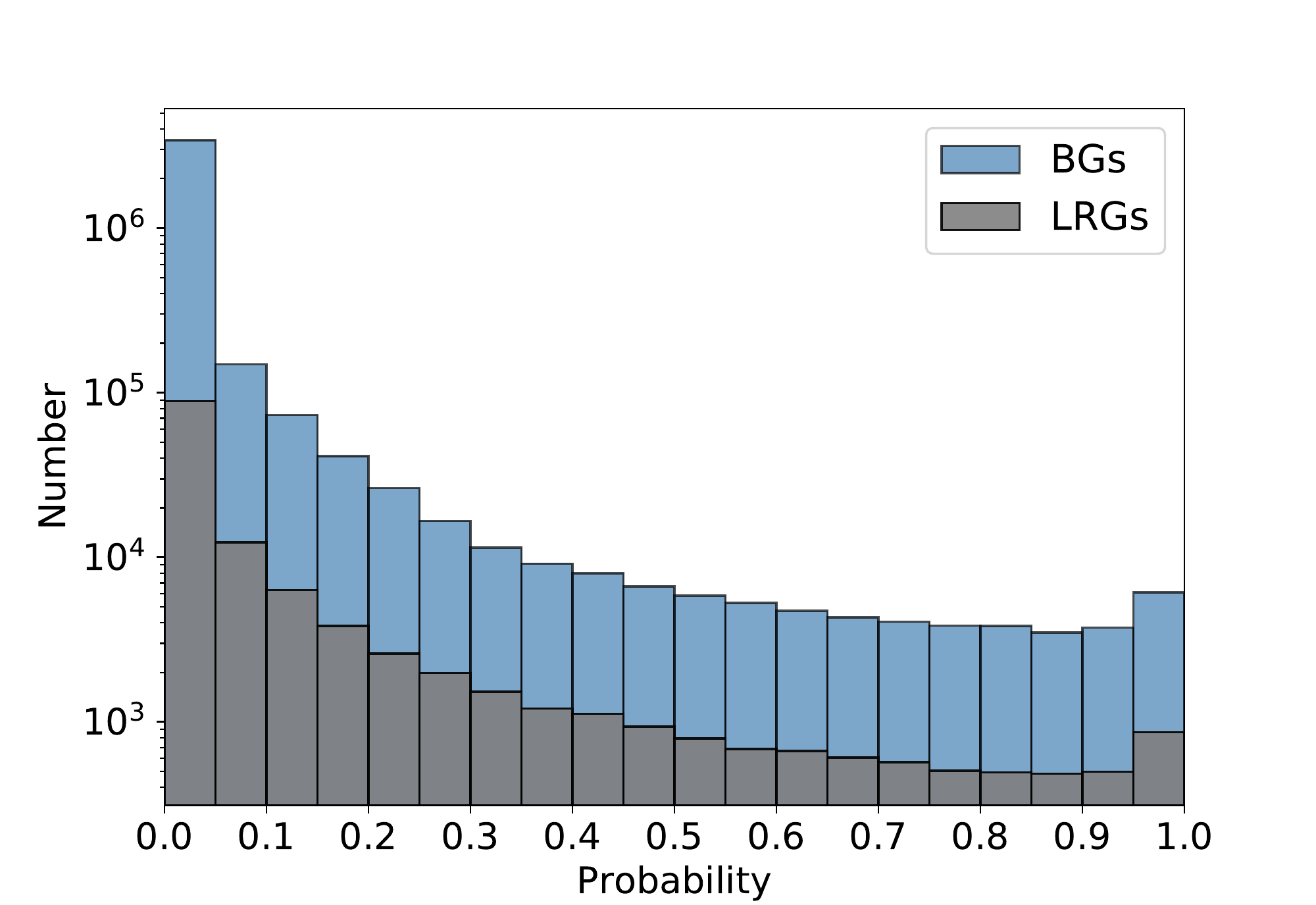}\vspace{3pt}
    \caption{The probability distribution of the predictive LRGs (grey) and BGs (blue). The two distributions follow the logarithm form, which are just as we have expected.}
    \label{fig:prob_distrib}
\end{figure}

In Fig.~\ref{fig:prob_distrib} we show the distributions of the CNN probability of LRG and BG samples. Overall, the two distributions follow a characteristic logarithmic trend, which is scaled according to the relative sample sizes. This trend is as expected, because targets with smaller CNN probabilities tend to be no-lens while those with larger probabilities tend to be real lenses and there are far more no-lenses in the real predictive samples compared to the lenses.

\subsection{Visual inspection}
\label{Vis_inspection}

Both lists of candidates (from LRG and BG samples) are definitely larger than the number of real lenses one can expect in the covered area \citep[$\sim 500$ in 1000 deg$^2$,][]{collett2015}, which means that these samples are dominated by false positives. 
In order to optimize the next visual inspection step, and give more time to inspectors to concentrate on significant candidates, we decided to have a first pass to filter clear false positives. 
In this case, only one observer had the task to inspect all candidates (2848 from LRG sample plus 3552 from BG sample) and excluded obvious non-lenses from the final sample to inspect. 
In this preliminary phase, we have also excluded all the lens candidates found by the CNN from P+19 and \cite{2017MNRAS.472.1129P}, including the LinKS sample, the bonus sample and any others they mentioned. The final number of candidates that survived this process was 286, 133 from the LRG sample and 153 from the BG sample.

\begin{figure}
    \centering
    \includegraphics[width=9cm]{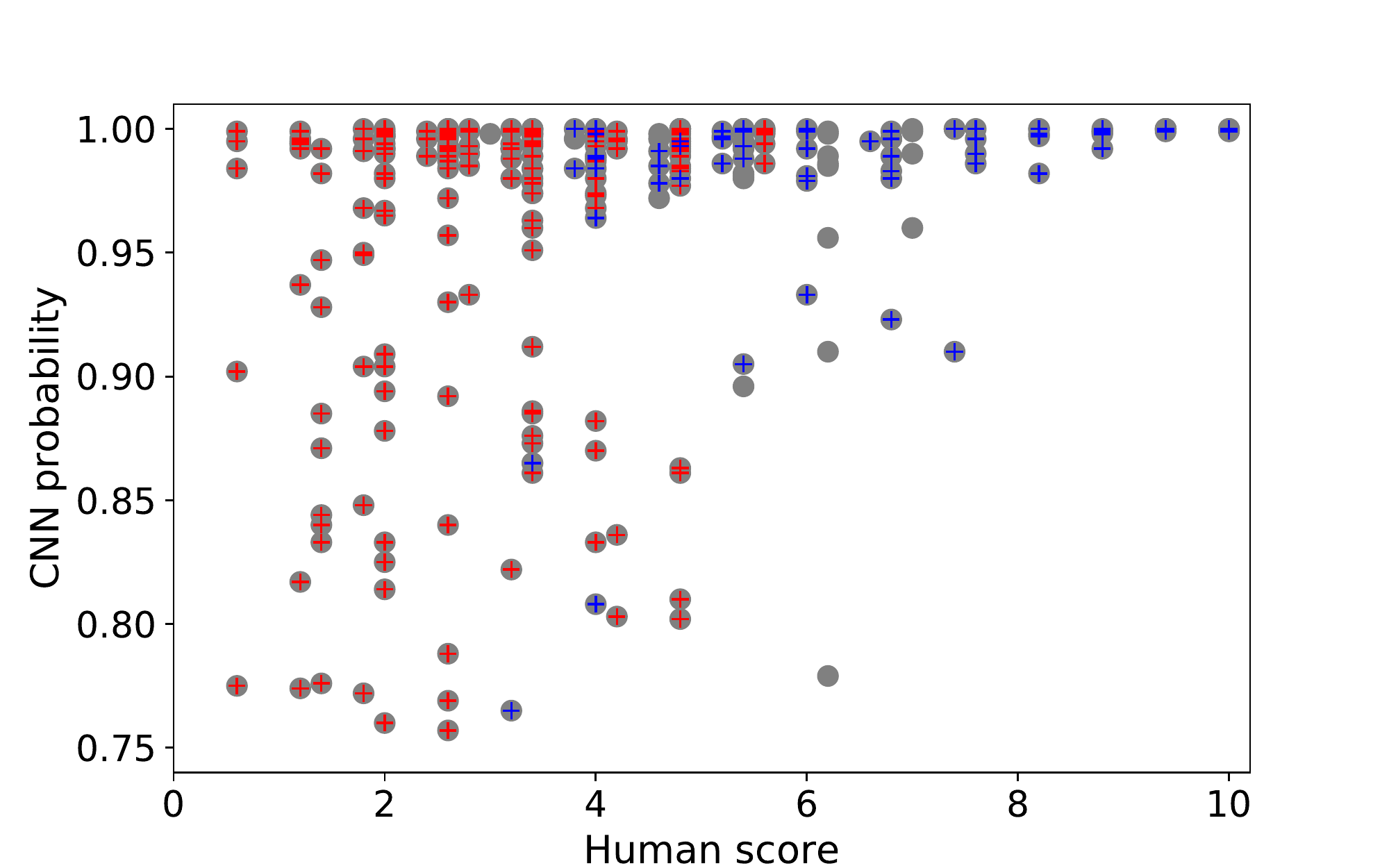}\vspace{3pt}
    \caption{CNN probability $p_{\rm cnn}$ against human probability $p_{\rm hum}$ for the 286 new candidates that passed both the ML and human thresholds. Points marked with blue crosses represent the systems for which at least one inspector gave a score of 10 (i.e. sure lens) while points marked with red crosses represent the systems for which at least one inspector gave a score equal to 0 (i.e. not a lens).}
    \label{fig:prob-score_HQlens}
\end{figure}

\begin{figure*}[htbp]
\centerline{\includegraphics[width=18cm]{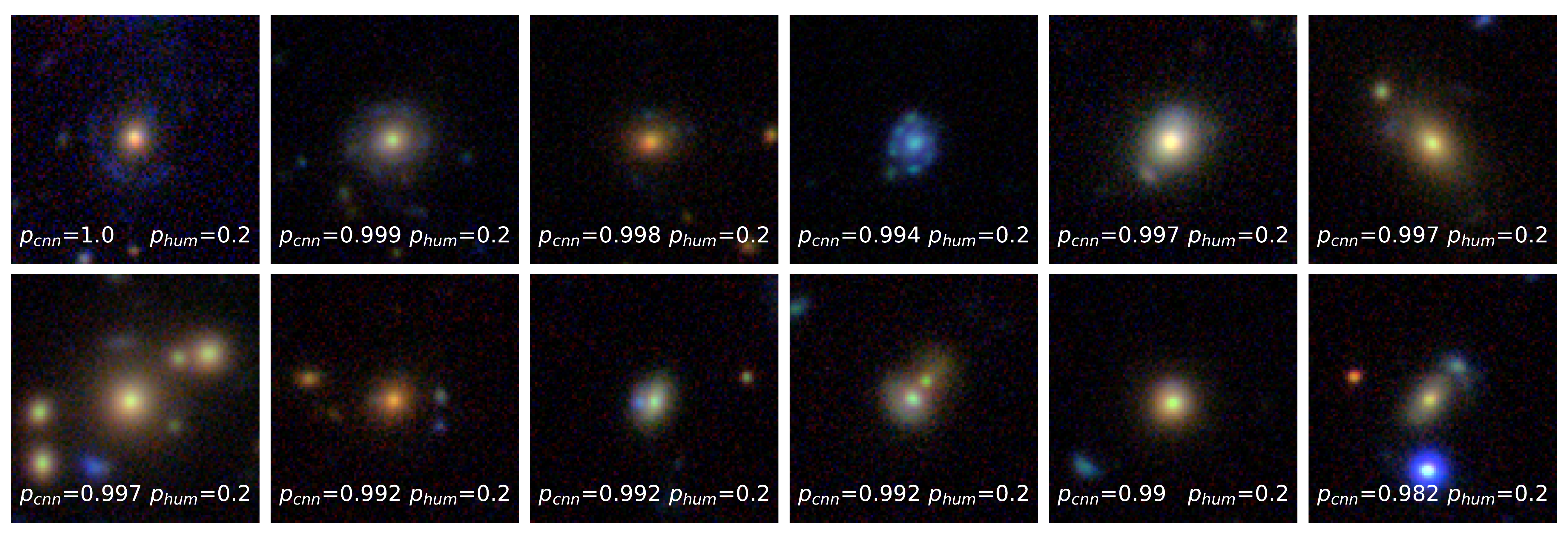}}
\caption{Candidates with high CNN probability ($p_{\rm cnn}\geq 0.97$) and low human score ($p_{\rm hum}=0.2$). There are some arc-like but not lens features (interactions, spiral arms, rings etc) that can give rise to some high $p_{\rm cnn}$}
\label{fig:high_prob_low_score}
\vspace{0.5cm}
\end{figure*}

The next step was to let five observers inspecting the objects selected on the basis of the CNN probability and that passed the visual pre-selection "cleaning". To this purpose we created color-cutout of $20''\times20''$, combining the g,r,i bands and let 5 people inspect the sample of pre-selected 286 objects in a blind way. The inspectors had to assign to each system a quality letter, following an $ABCD$ scheme where $A$ is a sure lens, $B$ is maybe a lens, $C$ is maybe not a lens and $D$ is not a lens, which we associated to a mark of 10, 7, 3, 0 respectively, to convert the quality flags into a score. 

We stress here that visual inspection does not provide a proof that a candidate lens is real. In this respect,  until we have access to a statistically large sample of spectroscopically confirmed lenses in KiDS, the ML will reproduce the human bias to define a lens as real. The best way to reduce this bias is indeed to increase the number of independent team members performing the visual assessment of the CNN lenses, as already stated in P+19. This is why in this paper, we always use five different inspectors to grade the candidates.

Finally, we defined a human probability as $p_{\rm hum}=s_{\rm ave}/10$ where $s_{\rm ave}$ is the average score from 5 inspectors. This human scoring returned 18 candidates with very high probability ($p_{\rm hum}\geq 0.8$) and another 10 with slightly lower probabilities ($0.7\leq p_{\rm hum}\leq0.8$) but still very convincing. These objects received all very high values also from the CNN as it can be seen in Fig. \ref{fig:prob-score_HQlens}. In this figure we plot the CNN probability $p_{\rm cnn}$ versus the human probability $p_{\rm hum}$. The 28  candidates are located in the top right corner of the plot, they have received both high probabilities from CNN and humans.

Moving toward to lower $p_{\rm hum}$, in the plot one should also expect the $p_{\rm cnn}$ to decrease and ideally the two quantities should be correlated. Instead, there is no clear correlation between the $p_{\rm cnn}$ and $p_{\rm hum}$, as the CNN gives a higher significance also to candidates that are poorly ranked by humans, although we observe a clear increase on the scatter between the two quantities. 
%the scatter increases too. 
In the upper left corner of Fig.~\ref{fig:prob-score_HQlens}, there are systems with very high $p_{\rm cnn}$($\geq 0.97$) but very low $p_{\rm hum}$($\leq0.4$). In these cases, either the CNN performs better than human eyes to detect real features that are not recognized by the inspectors, or the CNN more easily confuses features that can mimic gravitational arcs and {\it mupols}, which are more likely considered false positives from humans. 
Fig.~\ref{fig:high_prob_low_score} clearly demonstrates that the latter option is more likely the case. We show here a few cases of candidates with high $p_{\rm cnn}$($\geq 0.97$) and low $p_{\rm hum}$($=0.2$). Most of them are likely to be false positives since they show features (interactions, spiral arms, rings etc.) that mimic both faint arcs and {\it mupols}. This suggests that further effort is needed to improve the training set, by including more accurate ``negatives''. However, for these systems, we cannot rule out the possibility that small Einstein-radius lenses can be found by the CNN, but are harder to see by the inspectors, since the lensed images are hidden in the light of the central galaxies. But this problem can be partly overcome in the future by removing the light of the foreground galaxies.

In the middle region of Fig.~\ref{fig:prob-score_HQlens}, there are candidates for which the inspectors did not unanimously agreed on the classification and thus the final human probabilities are in the range of  $0.4 \leq p_{\rm cnn}\leq 0.6$. Here a large scatter in the CNN probability is found
probably because the machine tends to pick some features that have a lower SNR and are considered not totally convincing for humans. 

In order to figure how plausible the high $p_{\rm cnn}$ can be in this range of $p_{\rm hum}$, we marked all the points in Fig.~\ref{fig:prob-score_HQlens} for which at least one inspector considered the system as a sure lens (i.e. gave a grade of 10) with blue crosses. 
Many of these systems turned out to be {\it mupols}. This might indicate that the CNN has to be improved in the selection of this particular category lenses. 
We expect to qualify better these candidates with forthcoming experiments, training the
CNN on this specific class of systems. 
We note that red crosses indicate instead systems for which at least one inspector gave a score$=0$ (i.e. considered that object as a clear contaminant). 

\subsection{Ranking the candidates} 
Overall, Fig.~\ref{fig:prob-score_HQlens} suggests that, neither the $p_{\rm cnn}$ nor the $p_{\rm hum}$ are, alone, fully suitable parameters to rank the lenses (note that this is true for the current CNN, but might not be true for better networks). Hence, we decided to combine the two quantities to find a compromise between the CNN and human ``predictions'' and adopt a pseudo (joint) probability as a metric to rank the candidates:  
\begin{equation}
{\tt P}=p_{\rm cnn}*p_{\rm hum}   
\end{equation}
Using this probability, we identify 82 candidates with {\tt P}-value$\geq 0.5$, which we define high-quality lens candidates. Among them, 26 candidates represent a ``golden'' sample with {\tt P}-value $\geq0.7$, all showing 
obvious lens features and thus very suitable for spectroscopic follow-up observations.  

In Table \ref{tb:tb1}, we report the lens ID, the KiDS name, the  coordinates, the $r-$band magnitude, the photometric redshift, the average score $s_{\rm ave}$ from the inspectors, the $p_{\rm cnn}$ and {\tt P}-values of the 82 high-quality candidates, ranked in order of decreasing P-value.
Finally, in the last column of the table, we report the number of inspectors that gave a 0-score to that particular objects. 
In fact, as we described at the beginning of Section \ref{Vis_inspection}, a first pre-filtering of the 6400 objects with a $p_{\rm cnn}$ higher than the threshold was made by one single inspector. This person excluded obvious non-lenses from the final sample of candidates (286) that where then passed to other four people. 
This can be interpreted as assigning a 0-score to the excluded objects. Thus, formally, we should now exclude all systems where at least one of the remaining inspectors gave a 0-score. 
In this case, we would get rid of most of the low-scores and lower $p_{\rm cnn}$ in Fig.~\ref{fig:prob-score_HQlens} (in the bottom, left region of the plot), where we mark with red crosses systems that received at least one 0-score.
However, at the same time, we would also exclude many objects that received a very high grade from the CNN and could still be reliable candidates. We therefore decided to keep and flag these systems since, as already stressed, we have no way to understand if visual inspection works better/worse than CNN. We thus believe that reporting the number of inspectors that gave a zero on Table \ref{tb:tb1} and on the stamps we show in Fig. \ref{fig:high_quality_candidates}, is the best way to let readers judge by themselves.
 
\subsection{The high-quality lens sample}
\label{sec:HQ}
The 82 high-quality candidates, ranked in order of decreasing {\tt P}-value are shown in Fig.~\ref{fig:high_quality_candidates}.
%\ref{fig:high_quality_candidates1}. 
The stamps ($20''\times20''$) are obtained by combining $g$, $r$, and $i$ band images. We stress that the intrinsic signal-to-noise ratio (SNR) can change quite a lot in the different bands since $g$- and $i$-bands have worse seeing and depth with respect to the $r$-band. This might also be a factor of discrepancy between the CNN and human score, since the former only uses $r$-band while the visual inspection is made on the color-combined images and could be driven more by the combined SNR. We will expand the CNN predictions to other bands in forthcoming analyses \citep[see also the first attempt of this kind in][]{2019MNRAS.484.3879P,2019MNRAS.482..807P}. We stress that this implementation needs
very accurate color information when building the training sample. The colour distribution of the sources has to reproduce the realistic colours of real galaxies, otherwise its usage can lead to contradictory results. For instance, \cite{2019A&A...625A.119M} showed that, for their testing sample, multi-colour data are powerful in lensing search, as multi-band ground based data can reach better performance when compared with single-band space based data with lower noise and higher resolution. However, \cite{2019MNRAS.482..807P} also found that the color information can only partially help to improve the predictive ability of the CNN, since the CNN is mostly driven by morphology. The addition of color information might become troublesome if lens colors are heavily contaminated by the colors of the sources (i.e. looking bluer that they truly are, e.g., because of the close presence of very bright quasars). Thus a very careful identification of proper color-cuts and a proper training sample, reproducing the variety of colours and magnitudes of real lenses, are needed to make the multi-band approach effective.

At first glance, the majority of the candidates show distinguishable arc-like features, but some {\it mupols} candidates are also present. These candidates increase the number of previously found lensed quasar candidates in KiDS, using information from source colors in optical and infrared \citep[see e.g][]{2018MNRAS.480.1163S, 2019A&A...632A..56K, 2019MNRAS.484.3879P}. In particular, the ID=1 shows a very convincing peculiar Einstein cross configuration, while ID=12 seems to be a classical quadruplet in a fold-configuration. Also, ID=5 is likely a quad, with broad peaks due to the worse i-band seeing that shall be dominant, given the peculiar red color of the arc. These  objects are definitely very interesting for spectroscopic follow-up as, if confirmed, they will increase the number of know quads that are particularly useful for monitoring campaigns aimed at accurate measurements of the Hubble constant ($H_0$, \citealt{2017MNRAS.468.2590S}, \citealt{2019arXiv190704869W}).

Another important note is that about half of the candidates in the 'golden sample' are found in the BG sample (e.g. ID=3, 7, 10, 12, 13, 15, 16, 17, 18, 19, 20), which demonstrates that the ability of the CNN to find arcs and {\it mupols} around these systems has not been particularly affected by the training sample based on LRGs only (see Section\ref{sec:training}).

Finally, the CNN has captured some larger Einstein radii from group/clusters like ID=7 which shows a very faint and very red central deflector but a relatively large Einstein radius ($\sim5''$), with 3 arc-like images on the left and one point-like image on the right. The deflector has a high photo-z ($z_{\rm phot}=0.86$, the highest in the candidate list), which is coherent with the red color and the compact size. This is likely to be a dark matter rich system with one of the largest arc separation from an individual galaxy, especially considering the high redshift of the deflector. However, we can not exclude the possibility that this system is a galaxy group, since there at least three reddish objects in the vicinity of the lens galaxy candidate. If their redshifts are comparable with that of the central object, then this could be a lensing event from a small group, justify in this way the larger Einstein Radii. We have checked the photometric redshifts and this does not look to be the case. However we stress that the photometric redshifts are not always accurate.

The majority of the remaining high graded systems show quite regular arcs, and also pseudo-Einstein rings, like ID=25, 30, 33, 40, 47.

\begin{figure}
\vspace{-0.3cm}
    \centering
    \includegraphics[width=8.5cm]{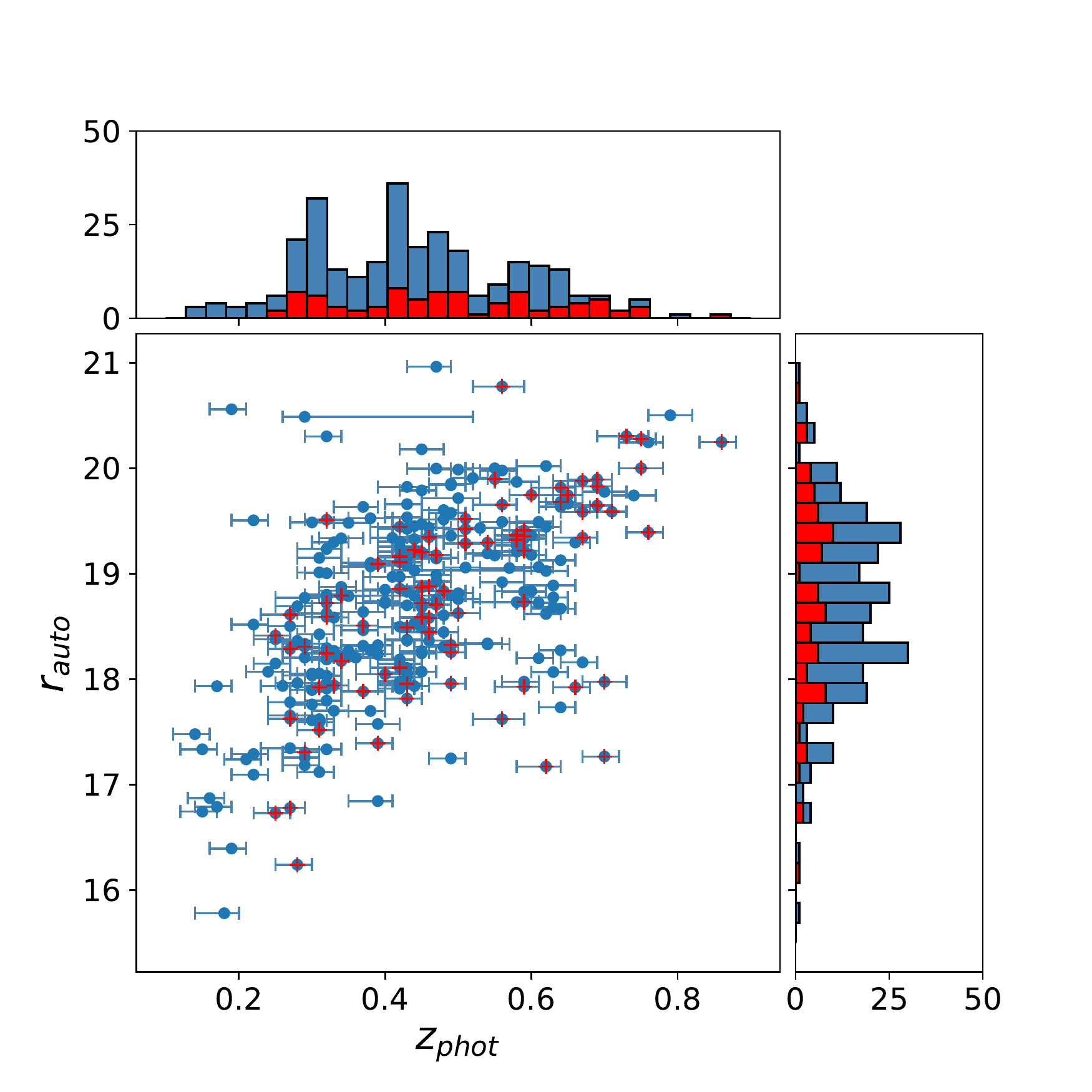}
    \caption{The distribution of the 286 lens candidates in the photometric redshift--luminosity space of the foreground deflectors. The dots marked by red crosses are the first 82 candidates shown in Fig.~\ref{fig:high_quality_candidates}. The error bars on the $r_{auto}$ magnitudes are smaller than the symbol sizes.}
    \label{fig:z_mag}
\end{figure}

In Fig.~\ref{fig:z_mag} we show the distribution of the deflectors in the photometric redshift--luminosity space. Photometric redshifts ($z_{\rm phot}$) are taken from the KiDS catalog and they have been obtained using BPZ \citep[for details, please see][]{2019A&A...625A...2K}. A correlation between the two quantities is clearly visible, as expected since, at fixed intrinsic luminosity (we remind the reader that we pre-selected bright galaxies only), the further a galaxy is (i.e. higher $z_{\rm phot}$ ) the smaller the apparent luminosity is. The correlation and the overall distribution in redshift and luminosity does not change if we include only the candidates in the top 82 ranking (marked by red crosses). The photo-z distribution is quite large in redshift and goes from $\sim0.2$ to $\sim0.8$. In addition, no correlation between the $z_{\rm phot}$ and the {\tt P}-value is found, as, for example, we have lenses with redshift $\sim0.2$ and $\sim0.8$ among the first 12 ranked candidates, and similarly in the second dozen in the ranking. In general, the redshift distribution of the new lens candidates seems slightly larger than the ones from P+19 that have almost no lenses above $z\sim0.6$. 

\section{Discussion}
\label{sec:disc}
The main aim of this paper is to report newly discovered high confidence strong lens candidates in the fourth KiDS data release, KiDS-DR4. These candidates have been found applying a CNN classifier that we recently developed following the prescription by P+19. 

The first question one might ask is what is the difference between the candidates from the two trained CNNs. 
The architecture and the depth of the network of the CNN we employ in this paper are identical to that of P+19. Hence the differences in number of candidates and mainly comes from the fact that here we expand both the predictive and the training samples. 
The second question is whether the complementarity of the two approaches can achieve the best completeness of the population of observable gravitational lens candidates. Finally a third question is if the number density of these lenses matches with expectations from simple statistical models (e.g. \citealt{2010MNRAS.405.2579O, collett2015}). This latter question is definitely relevant, but beyond the purpose of this paper as giving an answer would require a deeper analysis of the results coming from different methods. Possibly, this answer can come from an appropriate challenge comparing more techniques (not only the ones developed from our group) which should be run on the same (simulated) dataset, using different types of training samples or on different (real) predictive samples in order to establish if there is an optimal combination of methodologies to obtain the maximum possible completeness.

For the purpose of the current paper, we limit here to discuss four basic differences between the new CNN and the one from P+19. 
The first difference is the area coverage: in P+19 they missed $\sim100$ tiles that have made available for the final release and also they removed the masked regions ($\sim$100$-$200 sq.deg.) by setting ima$\_$flags$ = 0$ in all the 4 KiDS bands ($u$, $g$, $i$, $r$). In this work, we used all the 1006 publicly available tiles and did not remove the masked regions.
The second difference comes from the number of bands adopted: we used $r-$band only while P+19 has tested both 1 ($r$) and 3-bands ($g$, $i$, $r$). This does not necessarily impact the performance of our new CNN. In fact, the seeing in $g$ and $i$ band is in many case worse than that in $r-$band images. This could reduce the $P-$value returned by the $3-$bands based CNN.
A third relevant difference is the training sample. In fact, with respect to P+19, we extended the number of LRGs that we used to simulate real lenses, adding simulated arcs to them ({\tt positives}). Moreover, we also used $\sim7000$ more non-lensed galaxies to teach the CNN to exclude contaminants (see Section \ref{sec:training}). On the other side, we decided to only simulate 200\,000 mock lenses to training the CNN, while P+19 simulated 1\,000\,000. We did that because we checked that the addition of more mock lenses would not add more predictive power to the CNN.
The fourth difference is the dataset adopted to extract the predictive sample: P+19 have applied the CNN to KIDS DR4 {\it pre-published} data, while we have used the sample qualified for the ESO data release. As already mentioned, these two different datasets have different photometry parameters available (in the ESO DR4 the Krone-like magnitudes are available only for the r-band). This resulted in a different LRG sample (our selection included $\sim$126\,000 galaxies, while P+19 used $\sim$88\,000) mainly due to a different color definition (despite the same cuts adopted). In P+19 the colours are computed from the different bands {\tt mag$\_$auto}, while we use the {\tt COLOUR$\_$GAAP} columns given in the multi-band catalog and computed from {\tt MAG$\_$GAAP} magnitudes instead. 

This very qualitative comparison does not give a measure of the relative performances of the two CNNs, but possibly reveals their complementarity. 
As mentioned earlier a full comparison of the performances of the two networks is beyond the purposes of this paper, and we will discuss the differences in their detected systems in a future work. 

Other future developments will be oriented to implementing the completeness of the classifier and reduce the contamination from false positives. For instance, in Section \ref{sec:predictive_data} we have anticipated that a first implementation will consist in the training of the CNN classifier with $g, r, i$ color-composite images. 
%in the next work. 
We also plan to apply both the 1-band image trained CNN and 3-band image trained CNN to the future KiDS DR5, where we expect to substantially increase the number of final high-quality candidates in KiDS, since the total covered area will increase by $\sim 30$\%. 

In the future, our CNN will be easy adapted to the LSST as both the pixel scale ($0.2''$/pixel) and seeing ($<0.8''$) are very similar to the ones of KiDS. We will train the CNN on a simulated sample of lensed arcs and quasars built on LSST-like images (e.g. mock observations) in preparation for running the CNN on real LSST images to find real candidates. According to lens forecasts on LSST, we expect to collect $10^5$ lenses at the end of the full depth survey. In this respect, we expect to give a contribution to the ongoing effort to built the necessary machinery to get the completeness of the real lenses search close to 100\%. Apart from applying our CNN to LSST data, we also plan to apply CNN to CSST and EUCLID which will provide, from space, much better image quality, and %both 
expects to find also $\sim10^5$ new strong lenses.

\section{Conclusions}
We have developed a new CNN classifier to search for strong lens candidates in KiDS DR4, based on the prescription from a former CNN applied to the same KiDS DR4 by P+19. The new CNN makes use of  independent codes (both for the network and the simulated arcs) and different training and predictive samples. When applied to %the LRG sample, 
a sample of LRG as done in P+19, the new CNN classifier found 90\% of the high-quality candidates already presented in P+19, (including $10$ new ``golden'' lenses, ID=1, 2, 4, 5, 6, 8, 9, 11, 14, 21; see also below). Moreover, by applying this CNN classifier to the whole predictive dataset (not only LRG but also BG sample without any color cut applied), and combine this with human visual inspection, we found a total of 286 new lens candidates, including arcs, complete rings, but also multiple lensed images (e.g. Einstein Crosses and quadruplets).  We ranked the candidates by combining the CNN probability and the visual score P=$p_{\rm cnn}*p_{\rm hum}$, presented the parameters in Table~\ref{tb:tb1} and show the color-combined cutouts in Fig.~\ref{fig:high_quality_candidates} for the first 82 high-quality candidates with {\tt P}-value $\geq0.5$. Among them, 26 candidates, defined as ``golden sample'', have a very high probability to be real lenses and are suitable for follow-up observations. We finally provided a qualitative comparison between the CNN presented here and that presented in P+19, showing that the nets have comparable performances.  A quantitative, statistical and more complete comparison will be performed in a forthcoming publication. 
Moreover, in the future, we also plan to extend the CNN to new upcoming ground-based surveys (e.g. LSST) and space missions (CSST and EUCLID) to find a large number of
good strong lens candidates suitable for future spectroscopic confirmation follow-up programs.

\section*{Acknowledgements}
NRN and RL acknowledge financial support from the “One hundred top talent program of Sun Yat-sen University” grant N. 71000-18841229.  NRN also acknowledge support from the European Union Horizon 2020 research and innovation programme under the Marie Skodowska-Curie grant agreement n. 721463 to the SUNDIAL ITN network. RL acknowledge support from Guangdong Basic and Applied Basic Research Foundation 2019A1515110286. CT acknowledges funding from the INAF PRIN-SKA 2017 program 1.05.01.88.04. CS has received funding from the European Union’s Horizon 2020 research and innovation programme under the Marie Skłodowska-Curie actions grant agreement No 664931.

\begin{figure*}[htbp]
\centerline{\includegraphics[width=16cm]{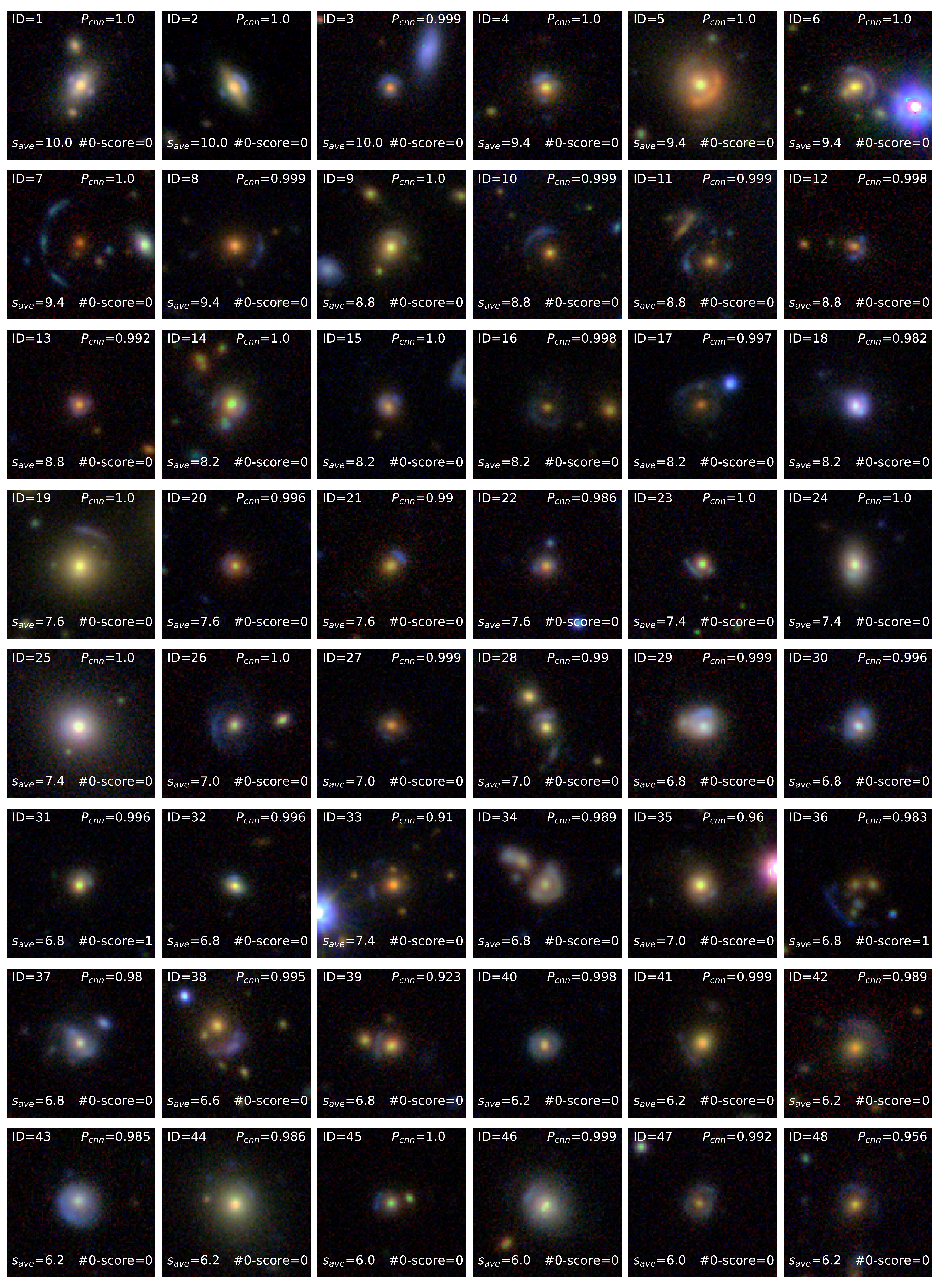}}
\caption{Colored stamps of the best 82 candidates, ranked according to the P-value. The stamps (20"x20") are obtained by combining g, r, and i KiDS images.}
%The first 100 candidates, ranked according to the {\tt P}-value. The stamps ($20''\times20''$) are obtained by combining $g$, $r$, and $i$ images.}
\label{fig:high_quality_candidates}
\vspace{0.5cm}
\end{figure*}

\addtocounter{figure}{-1}
\begin{figure*}[htbp]
\centerline{\includegraphics[width=16cm]{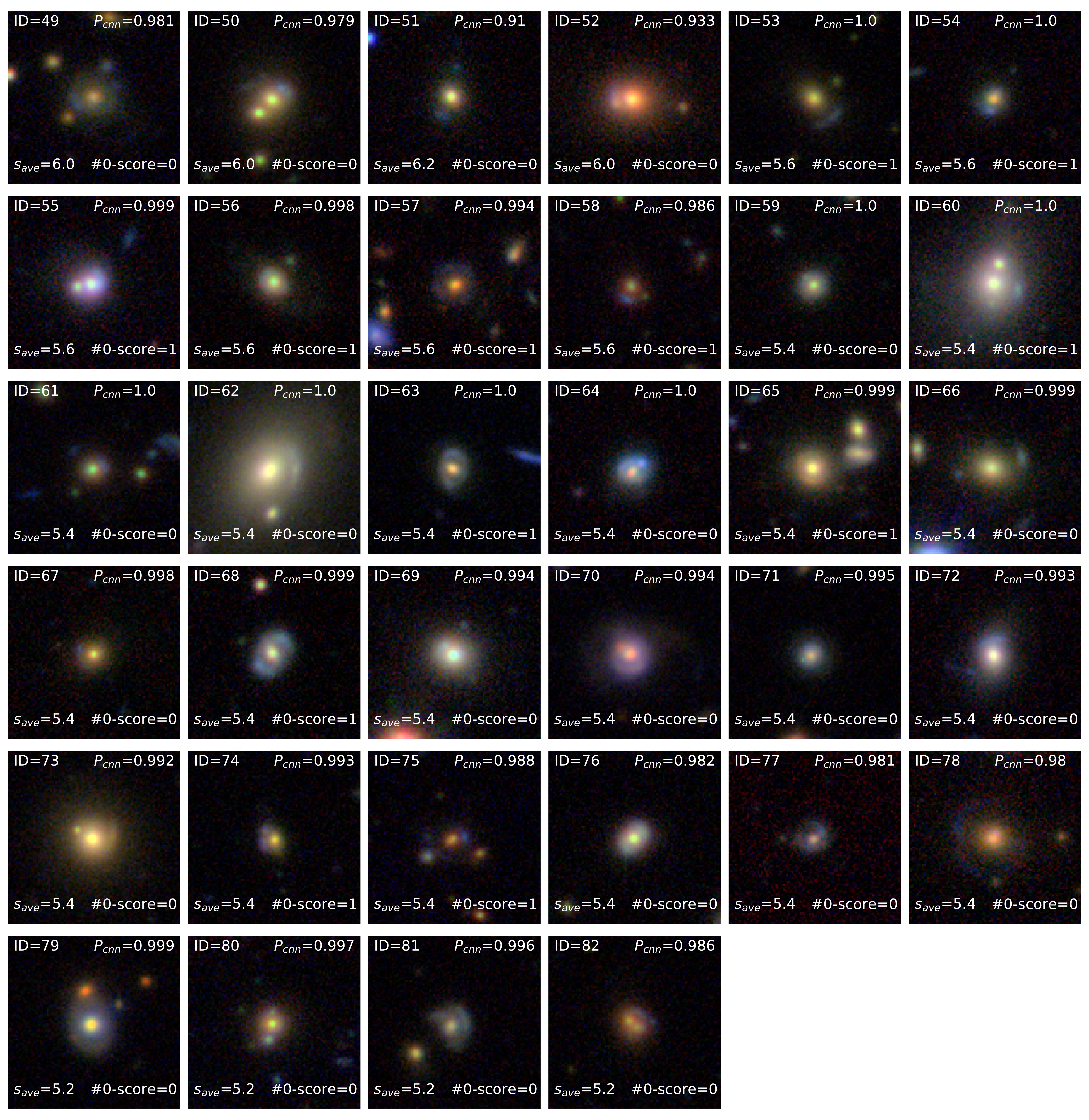}}
\caption{Continued}
\label{fig:high_quality_candidates1}
\vspace{0.5cm}
\end{figure*}

\begin{table*}[htbp]
\begin{center}
\caption{\label{tb:tb1} Properties of the best 82 lens candidates} %The properties of the sample.}
\begin{tabular}{l l l l l l l l l l l c }
\hline \hline
ID & KiDS$\_$NAME & RAJ2000 & DECJ2000 & $r_{auto}$ & $z_{phot}$ & $s_{\rm ave}$ & rms & $p_{\rm cnn}$ & {\tt P}-value & \#0-score\\
\hline
1 & KiDS J122456.016+005048.05 & 186.233401 & 0.846682 & 17.96 & $0.43^{+0.02}_{-0.04}$ & 10.0 & 0.0 & 1.0 & 1.0 & 0 \\
2 & KiDS J111253.976+001044.65 & 168.224904 & 0.179072 & 18.26 & $0.49^{+0.02}_{-0.03}$ & 10.0 & 0.0 & 1.0 & 1.0 & 0 \\
3 & KiDS J233533.673-322722.06 & 353.890307 & -32.456128 & 19.59 & $0.67^{+0.02}_{-0.03}$ & 10.0 & 0.0 & 0.999 & 0.999 & 0 \\
4 & KiDS J013425.700-295652.42 & 23.607086 & -29.947897 & 18.73 & $0.59^{+0.02}_{-0.04}$ & 9.4 & 1.2 & 1.0 & 0.94 & 0 \\
5 & KiDS J083933.372-014044.81 & 129.889052 & -1.679115 & 17.17 & $0.62^{+0.02}_{-0.04}$ & 9.4 & 1.2 & 1.0 & 0.94 & 0 \\
6 & KiDS J134032.074-003737.83 & 205.133643 & -0.627175 & 18.05 & $0.4^{+0.02}_{-0.04}$ & 9.4 & 1.2 & 1.0 & 0.94 & 0 \\
7 & KiDS J010704.918-312841.03 & 16.770493 & -31.478064 & 20.25 & $0.86^{+0.02}_{-0.03}$ & 9.4 & 1.2 & 1.0 & 0.94 & 0 \\
8 & KiDS J024228.926-294305.41 & 40.620528 & -29.718171 & 19.42 & $0.51^{+0.02}_{-0.03}$ & 9.4 & 1.2 & 0.999 & 0.939 & 0 \\
9 & KiDS J123554.179+005550.41 & 188.97575 & 0.93067 & 18.49 & $0.43^{+0.02}_{-0.04}$ & 8.8 & 1.47 & 1.0 & 0.88 & 0 \\
10 & KiDS J010606.232-310437.84 & 16.525969 & -31.07718 & 17.98 & $0.7^{+0.03}_{-0.03}$ & 8.8 & 1.47 & 0.999 & 0.879 & 0 \\
11 & KiDS J235728.351-352013.03 & 359.368133 & -35.336955 & 17.27 & $0.7^{+0.02}_{-0.03}$ & 8.8 & 1.47 & 0.999 & 0.879 & 0 \\
12 & KiDS J104223.359+001521.24 & 160.59733 & 0.2559 & 20.0 & $0.75^{+0.03}_{-0.03}$ & 8.8 & 1.47 & 0.998 & 0.878 & 0 \\
13 & KiDS J231242.301-332318.44 & 348.176257 & -33.388457 & 19.83 & $0.69^{+0.02}_{-0.04}$ & 8.8 & 1.47 & 0.992 & 0.873 & 0 \\
14 & KiDS J021504.013-284248.57 & 33.766723 & -28.713492 & 18.59 & $0.45^{+0.02}_{-0.03}$ & 8.2 & 1.47 & 1.0 & 0.82 & 0 \\
15 & KiDS J090507.336-001029.85 & 136.28057 & -0.17496 & 19.59 & $0.71^{+0.02}_{-0.04}$ & 8.2 & 1.47 & 1.0 & 0.82 & 0 \\
16 & KiDS J025334.181-284611.92 & 43.392423 & -28.769978 & 19.82 & $0.64^{+0.02}_{-0.04}$ & 8.2 & 1.47 & 0.998 & 0.818 & 0 \\
17 & KiDS J003151.142-312638.83 & 7.963094 & -31.44412 & 19.74 & $0.65^{+0.02}_{-0.04}$ & 8.2 & 1.47 & 0.997 & 0.818 & 0 \\
18 & KiDS J005540.416-290042.46 & 13.918401 & -29.011797 & 18.41 & $0.25^{+0.02}_{-0.03}$ & 8.2 & 1.47 & 0.982 & 0.805 & 0 \\
19 & KiDS J010257.486-291121.76 & 15.739527 & -29.189379 & 17.39 & $0.39^{+0.02}_{-0.03}$ & 7.6 & 1.2 & 1.0 & 0.76 & 0 \\
20 & KiDS J112900.041-014214.01 & 172.250173 & -1.703894 & 19.89 & $0.69^{+0.02}_{-0.04}$ & 7.6 & 1.2 & 0.996 & 0.757 & 0 \\
21 & KiDS J233620.351-352555.55 & 354.084799 & -35.4321 & 19.52 & $0.51^{+0.02}_{-0.04}$ & 7.6 & 1.2 & 0.99 & 0.752 & 0 \\
22 & KiDS J232152.835-275437.68 & 350.47015 & -27.910469 & 19.65 & $0.69^{+0.02}_{-0.03}$ & 7.6 & 1.2 & 0.986 & 0.749 & 0 \\
23 & KiDS J100108.387+024029.67 & 150.284948 & 2.67491 & 19.51 & $0.32^{+0.03}_{-0.03}$ & 7.4 & 2.58 & 1.0 & 0.74 & 0 \\
24 & KiDS J234338.567-335641.44 & 355.910697 & -33.944845 & 18.24 & $0.32^{+0.02}_{-0.03}$ & 7.4 & 2.58 & 1.0 & 0.74 & 0 \\
25 & KiDS J125834.900-004241.11 & 194.645418 & -0.711421 & 16.78 & $0.27^{+0.02}_{-0.03}$ & 7.4 & 2.58 & 1.0 & 0.74 & 0 \\
26 & KiDS J014518.788-290539.92 & 26.328284 & -29.094423 & 19.29 & $0.51^{+0.03}_{-0.03}$ & 7.0 & 0.0 & 1.0 & 0.7 & 0 \\
27 & KiDS J112152.078+023711.11 & 170.466993 & 2.619754 & 19.9 & $0.55^{+0.02}_{-0.04}$ & 7.0 & 0.0 & 0.999 & 0.699 & 0 \\
28 & KiDS J000820.374-342718.99 & 2.084894 & -34.455275 & 19.16 & $0.42^{+0.02}_{-0.03}$ & 7.0 & 0.0 & 0.99 & 0.693 & 0 \\
29 & KiDS J224258.953-351223.13 & 340.74564 & -35.206425 & 17.92 & $0.66^{+0.02}_{-0.03}$ & 6.8 & 2.23 & 0.999 & 0.679 & 0 \\
30 & KiDS J133317.497+005907.56 & 203.322906 & 0.985436 & 18.72 & $0.32^{+0.02}_{-0.03}$ & 6.8 & 2.23 & 0.996 & 0.677 & 0 \\
31 & KiDS J154712.516+002809.44 & 236.80215 & 0.469289 & 19.22 & $0.44^{+0.02}_{-0.03}$ & 6.8 & 3.65 & 0.996 & 0.677 & 1 \\
32 & KiDS J000517.478-352342.48 & 1.322827 & -35.395134 & 19.41 & $0.59^{+0.03}_{-0.03}$ & 6.8 & 2.23 & 0.996 & 0.677 & 0 \\
33 & KiDS J023714.701-280719.03 & 39.311257 & -28.121953 & 17.62 & $0.56^{+0.03}_{-0.04}$ & 7.4 & 2.58 & 0.91 & 0.673 & 0 \\
34 & KiDS J235920.307-290744.83 & 359.834614 & -29.129122 & 18.79 & $0.34^{+0.08}_{-0.03}$ & 6.8 & 2.23 & 0.989 & 0.673 & 0 \\
35 & KiDS J225409.348-274934.16 & 343.538954 & -27.826156 & 18.45 & $0.46^{+0.02}_{-0.04}$ & 7.0 & 0.0 & 0.96 & 0.672 & 0 \\
36 & KiDS J022956.259-311022.65 & 37.484416 & -31.172959 & 20.78 & $0.56^{+0.03}_{-0.04}$ & 6.8 & 3.65 & 0.983 & 0.668 & 1 \\
37 & KiDS J030628.054-291718.77 & 46.616892 & -29.288548 & 18.61 & $0.27^{+0.02}_{-0.04}$ & 6.8 & 2.23 & 0.98 & 0.666 & 0 \\
38 & KiDS J144950.559+005534.07 & 222.460665 & 0.926133 & 19.39 & $0.76^{+0.02}_{-0.03}$ & 6.6 & 3.14 & 0.995 & 0.657 & 0 \\
39 & KiDS J032230.223-344711.77 & 50.625931 & -34.786604 & 19.2 & $0.45^{+0.02}_{-0.03}$ & 6.8 & 2.23 & 0.923 & 0.628 & 0 \\
40 & KiDS J232911.441-324256.22 & 352.297671 & -32.715617 & 19.45 & $0.42^{+0.02}_{-0.04}$ & 6.2 & 1.6 & 0.998 & 0.619 & 0 \\
41 & KiDS J002105.099-283818.44 & 5.271248 & -28.638458 & 18.88 & $0.46^{+0.02}_{-0.04}$ & 6.2 & 1.6 & 0.999 & 0.619 & 0 \\
42 & KiDS J232039.461-281711.12 & 350.164421 & -28.286423 & 18.63 & $0.5^{+0.03}_{-0.03}$ & 6.2 & 1.6 & 0.989 & 0.613 & 0 \\
43 & KiDS J231310.384-344646.65 & 348.293267 & -34.779625 & 18.31 & $0.29^{+0.02}_{-0.03}$ & 6.2 & 1.6 & 0.985 & 0.611 & 0 \\
44 & KiDS J004439.128-291957.30 & 11.163036 & -29.332586 & 17.52 & $0.31^{+0.02}_{-0.03}$ & 6.2 & 1.6 & 0.986 & 0.611 & 0 \\
45 & KiDS J011731.429-314432.70 & 19.380956 & -31.742419 & 19.75 & $0.6^{+0.03}_{-0.03}$ & 6.0 & 2.68 & 1.0 & 0.6 & 0 \\
46 & KiDS J010649.164-284137.90 & 16.704852 & -28.693863 & 17.93 & $0.59^{+0.02}_{-0.04}$ & 6.0 & 2.68 & 0.999 & 0.599 & 0 \\
47 & KiDS J125814.219-005013.87 & 194.55925 & -0.837188 & 19.68 & $0.64^{+0.02}_{-0.03}$ & 6.0 & 2.68 & 0.992 & 0.595 & 0 \\
48 & KiDS J145325.778-003331.75 & 223.357411 & -0.558822 & 19.22 & $0.59^{+0.02}_{-0.04}$ & 6.2 & 1.6 & 0.956 & 0.593 & 0 \\
49 & KiDS J031142.084-341928.80 & 47.925354 & -34.324669 & 18.72 & $0.45^{+0.02}_{-0.04}$ & 6.0 & 2.68 & 0.981 & 0.589 & 0 \\
50 & KiDS J020554.272-342019.30 & 31.476136 & -34.338695 & 18.11 & $0.42^{+0.02}_{-0.04}$ & 6.0 & 2.68 & 0.979 & 0.587 & 0 \\
\hline
\end{tabular}
\end{center}
\textsc{      Note.} --- We list from Column 1 to 4, the ID, the KiDS name and the coordinates (in degrees) of the candidates, respectively. 
Column 5 lists the total magnitudes (rauto) obtained by from Sextractor. Column 6 lists the Photometric redshifts ($z_{\rm phot}$) taken from KiDS catalog,  using the BPZ code. Column 7 and 8 list the average scores from human inspection and the corresponding RMS. Column 9 list instead the probabily to be a lens from CNN. Column 10 then combines this information into the P-value threshold criterion defined in this work ($P=s_{\rm ave} \times p_{\rm cnn}/10$). Finally, Column 11 shows the numbers of  inspectors that gave a 0-score to that particular candidate (see text for more details).
%Column 1 are the ID of the candidates. Columns 2 are the KiDS name of the candidates. Columns 3 and 4 are the coordinates in RAJ2000 and DECJ2000, respectively. Column 5 are the total magnitude $r_{auto}$ parameters from Sextractor. Column 6 are the Photometric redshifts ($z_{\rm phot}$) from KiDS catalog, they are obtained using BPZ code. Column 7 are the average scores from human inspection while Column 8 are the corresponding RMS. Column 9 are the Probabilities from CNN. Column 10 are the criterion in this work defined as $s\times p _/10$. Column 11 are of numbers of the inspectors who given 0-score.
\end{table*}
\addtocounter{table}{-1}

\begin{table*}[htbp]
\begin{center}
\caption{\label{tb:tb1} %The properties of the sample.
Properties of the best 82 lens candidates}
\begin{tabular}{l l l l l l l l l l c }
\hline \hline
ID & KiDS$\_$NAME & RAJ2000 & DECJ2000 & $r_{auto}$ & $z_{phot}$ & $s_{\rm ave}$ & rms & $p_{\rm cnn}$ & {\tt P}-value & \#0-score\\
\hline
51 & KiDS J141913.862+025635.41 & 214.807762 & 2.94317 & 18.86 & $0.42^{+0.02}_{-0.03}$ & 6.2 & 1.6 & 0.91 & 0.564 & 0 \\
52 & KiDS J115110.395+025642.08 & 177.793313 & 2.945024 & 17.82 & $0.43^{+0.02}_{-0.03}$ & 6.0 & 2.68 & 0.933 & 0.56 & 0 \\
53 & KiDS J004558.739-331451.79 & 11.494746 & -33.24772 & 19.18 & $0.47^{+0.02}_{-0.03}$ & 5.6 & 2.8 & 1.0 & 0.56 & 1 \\
54 & KiDS J015928.393-330950.36 & 29.868305 & -33.16399 & 19.35 & $0.46^{+0.02}_{-0.04}$ & 5.6 & 2.8 & 1.0 & 0.56 & 1 \\
55 & KiDS J224712.244-333827.77 & 341.801017 & -33.641048 & 17.94 & $0.33^{+0.02}_{-0.04}$ & 5.6 & 2.8 & 0.999 & 0.559 & 1 \\
56 & KiDS J235255.478-291728.16 & 358.23116 & -29.291158 & 18.71 & $0.47^{+0.02}_{-0.03}$ & 5.6 & 2.8 & 0.998 & 0.559 & 1 \\
57 & KiDS J135138.926+002839.99 & 207.912195 & 0.477777 & 19.36 & $0.58^{+0.03}_{-0.03}$ & 5.6 & 2.8 & 0.994 & 0.557 & 1 \\
58 & KiDS J021609.168-293550.74 & 34.0382 & -29.597429 & 20.28 & $0.75^{+0.02}_{-0.03}$ & 5.6 & 2.8 & 0.986 & 0.552 & 1 \\
59 & KiDS J224308.305-344213.02 & 340.784606 & -34.703619 & 19.09 & $0.39^{+0.03}_{-0.04}$ & 5.4 & 1.96 & 1.0 & 0.54 & 0 \\
60 & KiDS J121234.927+000754.48 & 183.145531 & 0.1318 & 16.73 & $0.25^{+0.02}_{-0.03}$ & 5.4 & 3.5 & 1.0 & 0.54 & 1 \\
61 & KiDS J021555.605-342425.72 & 33.98169 & -34.407147 & 19.3 & $0.54^{+0.04}_{-0.04}$ & 5.4 & 1.96 & 1.0 & 0.54 & 0 \\
62 & KiDS J235510.007-283212.34 & 358.791698 & -28.536762 & 16.24 & $0.28^{+0.02}_{-0.03}$ & 5.4 & 1.96 & 1.0 & 0.54 & 0 \\
63 & KiDS J000012.031-310943.35 & 0.050133 & -31.162044 & 19.11 & $0.42^{+0.03}_{-0.03}$ & 5.4 & 3.5 & 1.0 & 0.54 & 1 \\
64 & KiDS J230527.508-313700.76 & 346.364619 & -31.61688 & 18.59 & $0.32^{+0.02}_{-0.03}$ & 5.4 & 1.96 & 1.0 & 0.54 & 0 \\
65 & KiDS J031516.618-310754.18 & 48.819245 & -31.131718 & 17.96 & $0.49^{+0.02}_{-0.03}$ & 5.4 & 3.5 & 0.999 & 0.539 & 1 \\
66 & KiDS J091113.492-000714.23 & 137.80622 & -0.12062 & 17.88 & $0.37^{+0.03}_{-0.03}$ & 5.4 & 1.96 & 0.999 & 0.539 & 0 \\
67 & KiDS J134455.641-002015.60 & 206.231838 & -0.337667 & 18.87 & $0.45^{+0.02}_{-0.03}$ & 5.4 & 1.96 & 0.998 & 0.539 & 0 \\
68 & KiDS J223123.786-282504.50 & 337.849109 & -28.417917 & 18.51 & $0.37^{+0.02}_{-0.04}$ & 5.4 & 3.5 & 0.999 & 0.539 & 1 \\
69 & KiDS J121319.575+014736.02 & 183.331564 & 1.793341 & 17.63 & $0.27^{+0.02}_{-0.03}$ & 5.4 & 1.96 & 0.994 & 0.537 & 0 \\
70 & KiDS J011045.486-290822.53 & 17.689526 & -29.139593 & 18.17 & $0.34^{+0.02}_{-0.03}$ & 5.4 & 1.96 & 0.994 & 0.537 & 0 \\
71 & KiDS J003242.839-310335.44 & 8.178496 & -31.059847 & 19.32 & $0.58^{+0.03}_{-0.03}$ & 5.4 & 1.96 & 0.995 & 0.537 & 0 \\
72 & KiDS J032426.994-290534.50 & 51.112476 & -29.092917 & 17.92 & $0.31^{+0.02}_{-0.03}$ & 5.4 & 1.96 & 0.993 & 0.536 & 0 \\
73 & KiDS J154051.806+010640.91 & 235.21586 & 1.111366 & 17.31 & $0.29^{+0.02}_{-0.03}$ & 5.4 & 1.96 & 0.992 & 0.536 & 0 \\
74 & KiDS J031609.185-340302.43 & 49.038271 & -34.050677 & 19.65 & $0.56^{+0.02}_{-0.04}$ & 5.4 & 3.5 & 0.993 & 0.536 & 1 \\
75 & KiDS J221400.330-292031.21 & 333.501378 & -29.342005 & 20.3 & $0.73^{+0.03}_{-0.04}$ & 5.4 & 3.5 & 0.988 & 0.534 & 1 \\
76 & KiDS J121314.238-001434.63 & 183.309326 & -0.242953 & 18.58 & $0.46^{+0.02}_{-0.04}$ & 5.4 & 1.96 & 0.982 & 0.53 & 0 \\
77 & KiDS J002141.664-301029.70 & 5.423603 & -30.174917 & 19.88 & $0.67^{+0.02}_{-0.04}$ & 5.4 & 1.96 & 0.981 & 0.53 & 0 \\
78 & KiDS J122335.140-021030.63 & 185.896418 & -2.175176 & 18.32 & $0.49^{+0.02}_{-0.03}$ & 5.4 & 1.96 & 0.98 & 0.529 & 0 \\
79 & KiDS J130115.900+025240.95 & 195.316253 & 2.878043 & 18.29 & $0.27^{+0.03}_{-0.03}$ & 5.2 & 2.86 & 0.999 & 0.519 & 0 \\
80 & KiDS J001810.363-285609.54 & 4.54318 & -28.935984 & 18.84 & $0.48^{+0.03}_{-0.03}$ & 5.2 & 2.86 & 0.997 & 0.518 & 0 \\
81 & KiDS J025717.233-271712.02 & 44.321807 & -27.286673 & 19.35 & $0.59^{+0.02}_{-0.03}$ & 5.2 & 2.86 & 0.996 & 0.518 & 0 \\
82 & KiDS J104119.501-000416.30 & 160.331257 & -0.071195 & 19.34 & $0.67^{+0.02}_{-0.04}$ & 5.2 & 2.86 & 0.986 & 0.513 & 0 \\
\hline 
\end{tabular}
\end{center}
\end{table*}


\begin{thebibliography}

\bibitem[Abadi et al.(2016)]{2016arXiv160304467A} Abadi, M., Agarwal, A., Barham, P., et al.\ 2016, arXiv e-prints, arXiv:1603.04467

\bibitem[Agnello et al.(2015)]{2015MNRAS.448.1446A} Agnello, A., Kelly, B.~C., Treu, T., et al.\ 2015, \mnras, 448, 1446

\bibitem[Agnello et al.(2018)]{2018MNRAS.479.4345A} Agnello, A., Lin, H., Kuropatkin, N., et al.\ 2018, \mnras, 479, 4345

\bibitem[Agnello \& Spiniello(2019)]{2019MNRAS.489.2525A} Agnello, A., \& Spiniello, C.\ 2019, \mnras, 489, 2525

\bibitem[Agnello et al.(2018a)]{2018MNRAS.479.4345A} Agnello, A., Lin, H., Kuropatkin, N., et al.\ 2018, \mnras, 479, 4345
\bibitem[Agnello(2018b)]{2018RNAAS...2...42A} Agnello, A.\ 2018, Research Notes of the American Astronomical Society, 2, 42

\bibitem[ALMA Partnership et al.(2015)]{2015ApJ...808L...4A} ALMA Partnership, Vlahakis, C., Hunter, T.~R., et al.\ 2015, \apjl, 808, L4

\bibitem[Amendola et al.(2018)]{2018LRR....21....2A} Amendola, L., Appleby, S., Avgoustidis, A., et al.\ 2018, Living Reviews in Relativity, 21, 2

\bibitem[Auger et al.(2009)]{2009ApJ...705.1099A} Auger, M.~W., Treu, T., Bolton, A.~S., et al.\ 2009, \apj, 705, 1099

\bibitem[Auger et al.(2010)]{2010ApJ...724..511A} Auger, M.~W., Treu, T., Bolton, A.~S., et al.\ 2010, \apj, 724, 511

\bibitem[Barnab{\`e} et al.(2012)]{2012MNRAS.423.1073B} Barnab{\`e}, M., Dutton, A.~A., Marshall, P.~J., et al.\ 2012, \mnras, 423, 1073

\bibitem[Bertin, \& Arnouts(1996)]{1996A&AS..117..393B} Bertin, E., \& Arnouts, S.\ 1996, \aaps, 117, 393

\bibitem[Blandford \& Narayan(1992)]{1992ARA&A..30..311B} Blandford, R.~D., \& Narayan, R.\ 1992, \araa, 30, 311

\bibitem[Bolton et al.(2008)]{2008ApJ...682..964B} Bolton, A.~S., Burles, S., Koopmans, L.~V.~E., et al.\ 2008, \apj, 682, 964

\bibitem[Bolton et al.(2006)]{2006ApJ...638..703B} Bolton, A.~S., Burles, S., Koopmans, L.~V.~E., et al.\ 2006, \apj, 638, 703  

\bibitem[Bolton et al.(2012)]{2012ApJ...757...82B} Bolton, A.~S., Brownstein, J.~R., Kochanek, C.~S., et al.\ 2012, \apj, 757, 82

\bibitem[Bonvin et al.(2017)]{2017MNRAS.465.4914B} Bonvin, V., Courbin, F., Suyu, S.~H., et al.\ 2017, \mnras, 465, 4914

\bibitem[Brownstein et al.(2012)]{2012ApJ...744...41B} Brownstein, J.~R., Bolton, A.~S., Schlegel, D.~J., et al.\ 2012, \apj, 744, 41

\bibitem[Cao et al.(2017)]{2017ApJ...835...92C} Cao, S., Li, X., Biesiada, M., et al.\ 2017, \apj, 835, 92

\bibitem[Chen et al.(2019)]{2019ApJ...881....8C} Chen, W., Kelly, P.~L., Diego, J.~M., et al.\ 2019, \apj, 881, 8

\bibitem[Claeyssens et al.(2019)]{2019MNRAS.489.5022C} Claeyssens, A., Richard, J., Blaizot, J., et al.\ 2019, \mnras, 489, 5022

\bibitem[Closson Ferguson et al.(2009)]{2009AAS...21346007C} Closson Ferguson, H., Armus, L., Borne, K., et al.\ 2009, American Astronomical Society Meeting Abstracts \#213 213, 460.07

\bibitem[Collett(2015)]{collett2015} Collett, T.~E.\ 2015, \apj, 811, 20

\bibitem[Collett et al.(2018)]{2018Sci...360.1342C} Collett, T.~E., Oldham, L.~J., Smith, R.~J., et al.\ 2018, Science, 360, 1342

\bibitem[Congdon \& Keeton(2018)]{2018pgl..book.....C} Congdon, A.~B., \& Keeton, C.\ 2018, Springer International Publishing

\bibitem[Cornachione et al.(2018)]{2018ApJ...853..148C} Cornachione, M.~A., Bolton, A.~S., Shu, Y., et al.\ 2018, \apj, 853, 148

\bibitem[de Jong et al.(2013)]{2013Msngr.154...44D} de Jong, J.~T.~A., Kuijken, K., Applegate, D., et al.\ 2013, The Messenger, 154, 44

\bibitem[Dobler et al.(2008)]{2008ApJ...685...57D} Dobler, G., Keeton, C.~R., Bolton, A.~S., et al.\ 2008, \apj, 685, 57

\bibitem[Eisenstein et al.(2001)]{2001AJ....122.2267E} Eisenstein, D.~J., Annis, J., Gunn, J.~E., et al.\ 2001, \aj, 122, 2267

\bibitem[Fukugita et al.(1992)]{1992ApJ...393....3F} Fukugita, M., Futamase, T., Kasai, M., et al.\ 1992, \apj, 393, 3

\bibitem[Gilman et al.(2018)]{2018MNRAS.481..819G} Gilman, D., Birrer, S., Treu, T., et al.\ 2018, \mnras, 481, 819

\bibitem[Hartley et al.(2017)]{2017MNRAS.471.3378H} Hartley, P., Flamary, R., Jackson, N., et al.\ 2017, \mnras, 471, 3378

\bibitem[He et al.(2015)]{2015arXiv151203385H} He, K., Zhang, X., Ren, S., et al.\ 2015, arXiv e-prints, arXiv:1512.03385

\bibitem[Hsueh et al.(2020)]{2020MNRAS.492.3047H} Hsueh, J.-W., Enzi, W., Vegetti, S., et al.\ 2020, \mnras, 492, 3047

\bibitem[Huang et al.(2016)]{2016arXiv160806993H} Huang, G., Liu, Z., van der Maaten, L., et al.\ 2016, arXiv e-prints, arXiv:1608.06993

\bibitem[Ivezi{\'c} et al.(2014)]{2014sdmm.book.....I} Ivezi{\'c}, {\v{Z}}., Connelly, A.~J., VanderPlas, J.~T., et al.\ 2014, Statistics

\bibitem[Jacobs et al.(2017)]{2017MNRAS.471..167J} Jacobs, C., Glazebrook, K., Collett, T., et al.\ 2017, \mnras, 471, 167

\bibitem[Jacobs et al.(2019)]{2019ApJS..243...17J} Jacobs, C., Collett, T., Glazebrook, K., et al.\ 2019, \apjs, 243, 17

\bibitem[Keeton(1998)]{1998PhDT.........6K} Keeton, C.~R.\ 1998, Ph.D. Thesis

\bibitem[Khramtsov et al.(2019)]{2019A&A...632A..56K} Khramtsov, V., Sergeyev, A., Spiniello, C., et al.\ 2019, \aap, 632, A56

\bibitem[Kochanek(2019)]{2019arXiv191105083K} Kochanek, C.~S.\ 2019, arXiv e-prints, arXiv:1911.05083

\bibitem[Koopmans \& Treu(2003)]{2003ApJ...583..606K} Koopmans, L.~V.~E., \& Treu, T.\ 2003, \apj, 583, 606 

\bibitem[Koopmans et al.(2006)]{2006ApJ...649..599K} Koopmans, L.~V.~E., Treu, T., Bolton, A.~S., Burles, S., \& Moustakas, L.~A.\ 2006, \apj, 649, 599

\bibitem[Koopmans et al.(2009)]{2009ApJ...703L..51K} Koopmans, L.~V.~E., Bolton, A., Treu, T., et al.\ 2009, \apjl, 703, L51

\bibitem[Kouw \& Loog(2018)]{2018arXiv181211806K} Kouw, W.~M., \& Loog, M.\ 2018, arXiv e-prints, arXiv:1812.11806

\bibitem[Krizhevsky et al.(2012)]{2012ANIPS} Krizhevsky, A.,  Sutskever,I. and G. Advances, Hinton. in neural information processing systems, page 1097--1105. (2012)

\bibitem[Kuijken et al.(2019)]{2019A&A...625A...2K} Kuijken, K., Heymans, C., Dvornik, A., et al.\ 2019, \aap, 625, A2

\bibitem[La Barbera et al.(2008)]{2008PASP..120..681L} La Barbera, F., de Carvalho, R.~R., Kohl-Moreira, J.~L., et al.\ 2008, \pasp, 120, 681

\bibitem[Lecun et al.(1998)]{1998IEEE..86..11L} Lecun, Y.,  Bottou, L., Bengio, Y. and Haffner, P., \ 1998,  \ Proceedings of the IEEE, 86, 11

\bibitem[Lemon et al.(2020)]{2020MNRAS.494.3491L} Lemon, C., Auger, M.~W., McMahon, R., et al.\ 2020, \mnras, 494, 3491

\bibitem[Li et al.(2017)]{2017MNRAS.468.1426L} Li, R., Frenk, C.~S., Cole, S., et al.\ 2017, \mnras, 468, 1426

\bibitem[Li et al.(2018)]{2018MNRAS.480..431L} Li, R., Shu, Y., \& Wang, J.\ 2018, \mnras, 480, 431 

\bibitem[Li et al.(2019)]{2019MNRAS.482..313L} Li, R., Shu, Y., Su, J., et al.\ 2019, \mnras, 482, 313

\bibitem[Marshall et al.(2016)]{2016MNRAS.455.1171M} Marshall, P.~J., Verma, A., More, A., et al.\ 2016, \mnras, 455, 1171

\bibitem[Metcalf et al.(2019)]{2019A&A...625A.119M} Metcalf, R.~B., Meneghetti, M., Avestruz, C., et al.\ 2019, \aap, 625, A119

\bibitem[Michalski(1986)]{1986mlaa.book.....M} Michalski, R.~S.\ 1986, Los Altos: M. Kaufmann Publication

\bibitem[Miyazaki et al.(2012)]{2012SPIE.8446E..0ZM} Miyazaki, S., Komiyama, Y., Nakaya, H., et al.\ 2012, \procspie, 84460Z

\bibitem[More et al.(2016)]{2016MNRAS.455.1191M} More, A., Verma, A., Marshall, P.~J., et al.\ 2016, \mnras, 455, 1191

\bibitem[Moster et al.(2010)]{2010ApJ...710..903M} Moster, B.~P., Somerville, R.~S., Maulbetsch, C., et al.\ 2010, \apj, 710, 903

\bibitem[Napolitano et al.(2016)]{2016ASSP...42..129N} Napolitano, N.~R., Covone, G., Roy, N., et al.\ 2016, The Universe of Digital Sky Surveys, 129

\bibitem[Nightingale et al.(2019)]{2019MNRAS.489.2049N} Nightingale, J.~W., Massey, R.~J., Harvey, D.~R., et al.\ 2019, \mnras, 489, 2049

\bibitem[Oguri \& Marshall(2010)]{2010MNRAS.405.2579O} Oguri, M., \& Marshall, P.~J.\ 2010, \mnras, 405, 2579

\bibitem[Petrillo et al.(2017)]{2017MNRAS.472.1129P} Petrillo, C.~E., Tortora, C., Chatterjee, S., et al.\ 2017, \mnras, 472, 1129

\bibitem[Petrillo et al.(2019a)]{2019MNRAS.484.3879P} Petrillo, C.~E., Tortora, C., Vernardos, G., et al.\ 2019, \mnras, 484, 3879

\bibitem[Petrillo et al.(2019b)]{2019MNRAS.482..807P} Petrillo, C.~E., Tortora, C., Chatterjee, S., et al.\ 2019, \mnras, 482, 807

\bibitem[Pourrahmani et al.(2018)]{2018ApJ...856...68P} Pourrahmani, M., Nayyeri, H., \& Cooray, A.\ 2018, \apj, 856, 68

\bibitem[Rawat et al.(2018)]{2017NC...29...1R} Rawat, Waseem \& Wang, Zenghui. (2017). Deep Convolutional Neural Networks for Image Classification: A Comprehensive Review. Neural Computation. 29. 1-98

\bibitem[Refsdal(1964)]{1964MNRAS.128..307R} Refsdal, S.\ 1964, \mnras, 128, 307

\bibitem[Roy et al.(2018)]{2018MNRAS.480.1057R} Roy, N., Napolitano, N.~R., La Barbera, F., et al.\ 2018, \mnras, 480, 1057

\bibitem[Ruff et al.(2011)]{2011ApJ...727...96R} Ruff, A.~J., Gavazzi, R., Marshall, P.~J., et al.\ 2011, \apj, 727, 96

\bibitem[Rydberg et al.(2019)]{2019MNRAS.tmp.2847R} Rydberg, C.-E., Whalen, D.~J., Maturi, M., et al.\ 2019, \mnras, 2847

\bibitem[Schneider et al.(1992)]{1992grle.book.....S} Schneider, P., Ehlers, J., \& Falco, E.~E.\ 1992, Gravitational Lenses

\bibitem[Schuldt et al.(2019)]{2019A&A...631A..40S} Schuldt, S., Chiriv{\`\i}, G., Suyu, S.~H., et al.\ 2019, \aap, 631, A40

\bibitem[Schwab et al.(2010)]{2010ApJ...708..750S} Schwab, J., Bolton, A.~S., \& Rappaport, S.~A.\ 2010, \apj, 708, 750

\bibitem[Seidel, \& Bartelmann(2007)]{2007A&A...472..341S} Seidel, G., \& Bartelmann, M.\ 2007, \aap, 472, 341

\bibitem[Sluse et al.(2019)]{2019MNRAS.490..613S} Sluse, D., Rusu, C.~E., Fassnacht, C.~D., et al.\ 2019, \mnras, 490, 613

\bibitem[Shu et al.(2015)]{2015ApJ...803...71S} Shu, Y., Bolton, A.~S., Brownstein, J.~R., et al.\ 2015, \apj, 803, 71

\bibitem[Shu et al.(2016a)]{2016ApJ...824...86S} Shu, Y., Bolton, A.~S., Kochanek, C.~S., et al.\ 2016, \apj, 824, 86

\bibitem[Shu et al.(2016b)]{2016ApJ...833..264S} Shu, Y., Bolton, A.~S., Mao, S., et al.\ 2016, \apj, 833, 264

\bibitem[Shu et al.(2017)]{2017ApJ...851...48S} Shu, Y., Brownstein, J.~R., Bolton, A.~S., et al.\ 2017, \apj, 851, 48

\bibitem[Sonnenfeld et al.(2013)]{2013ApJ...777...98S} Sonnenfeld, A., Treu, T., Gavazzi, R., et al.\ 2013, \apj, 777, 98

\bibitem[Sonnenfeld et al.(2020)]{2020arXiv200400634S} Sonnenfeld, A., Verma, A., More, A., et al.\ 2020, arXiv e-prints, arXiv:2004.00634

\bibitem[Spiniello et al.(2011)]{2011MNRAS.417.3000S} Spiniello, C., Koopmans, L.~V.~E., Trager, S.~C., et al.\ 2011, \mnras, 417, 3000

\bibitem[Spiniello et al.(2018)]{2018MNRAS.480.1163S} Spiniello, C., Agnello, A., Napolitano, N.~R., et al.\ 2018, \mnras, 480, 1163

\bibitem[Spiniello et al.(2019)]{2019MNRAS.483.3888S} Spiniello, C., Agnello, A., Sergeyev, A.~V., et al.\ 2019, \mnras, 483, 3888

\bibitem[Suyu et al.(2013)]{2013ApJ...766...70S} Suyu, S.~H., Auger, M.~W., Hilbert, S., et al.\ 2013, \apj, 766, 70

\bibitem[Suyu et al.(2017)]{2017MNRAS.468.2590S} Suyu, S.~H., Bonvin, V., Courbin, F., et al.\ 2017, \mnras, 468, 2590

\bibitem[The Dark Energy Survey Collaboration(2005)]{2005astro.ph.10346T} The Dark Energy Survey Collaboration\ 2005, arXiv e-prints, astro-ph/0510346

\bibitem[Tortora et al.(2010)]{2010ApJ...721L...1T} Tortora, C., Napolitano, N.~R., Romanowsky, A.~J., et al.\ 2010, \apjl, 721, L1

\bibitem[Treu \& Koopmans(2004)]{2004ApJ...611..739T} Treu, T., \& Koopmans, L.~V.~E.\ 2004, \apj, 611, 739

\bibitem[Treu et al.(2011)]{2011MNRAS.417.1601T} Treu, T., Dutton, A.~A., Auger, M.~W., et al.\ 2011, \mnras, 417, 1601

\bibitem[Treu \& SWELLS Team(2012)]{2012AAS...21931106T} Treu, T., \& SWELLS Team\ 2012, American Astronomical Society Meeting Abstracts \#219 219, 311.06

\bibitem[Turner et al.(1984)]{1984ApJ...284....1T} Turner, E.~L., Ostriker, J.~P., \& Gott, J.~R.\ 1984, \apj, 284, 1

\bibitem[Vegetti et al.(2012)]{2012Natur.481..341V} Vegetti, S., Lagattuta, D.~J., McKean, J.~P., et al.\ 2012, \nat, 481, 341

\bibitem[Weinberg et al.(2013)]{2013PhR...530...87W} Weinberg, D.~H., Mortonson, M.~J., Eisenstein, D.~J., et al.\ 2013, \physrep, 530, 87

\bibitem[Wong et al.(2013)]{2013ApJ...769...52W} Wong, K.~C., Zabludoff, A.~I., Ammons, S.~M., et al.\ 2013, \apj, 769, 52

\bibitem[Wong et al.(2019)]{2019arXiv190704869W} Wong, K.~C., Suyu, S.~H., Chen, G.~C.-F., et al.\ 2019, arXiv e-prints, arXiv:1907.04869

\bibitem[York et al.(2000)]{2000AJ....120.1579Y} York, D.~G., Adelman, J., Anderson, J.~E., et al.\ 2000, \aj, 120, 1579

\bibitem[Zhan(2018)]{2018cosp...42E3821Z} Zhan, H.\ 2018, 42nd COSPAR Scientific Assembly, E1.16-4-18

\bibitem[Zwicky(1937)]{1937PhRv...51..290Z} Zwicky, F.\ 1937, Physical Review, 51, 290

\end{thebibliography}
\end{document}